%% file: cas-dc-template.tex
\documentclass[a4paper,fleqn]{cas-dc}

\usepackage[numbers, sort&compress]{natbib}
\usepackage{subcaption}
\usepackage{svg}

\usepackage{geometry}
\usepackage{tabularx}
\usepackage{booktabs}
\usepackage[table]{xcolor}
\usepackage{amsmath}
\usepackage{xcolor}
\usepackage{soul}
\usepackage{float}
\sethlcolor{red!30} % sets highlight to light red

\usepackage{caption}   % per \captionof
\definecolor{lightblue}{RGB}{31,119,180}  % azzurro "matplotlib blue", elegante e leggibile

\def\tsc#1{\csdef{#1}{\textsc{\lowercase{#1}}\xspace}}
\tsc{WGM}
\tsc{QE}
\tsc{EP}
\tsc{PMS}
\tsc{BEC}
\tsc{DE}
\begin{document}
\let\WriteBookmarks\relax
\def\floatpagepagefraction{1}
\def\textpagefraction{.001}

\shorttitle{Toward Open ISA Supercomputing}

\shortauthors{E. Venieri et~al.}

\title [mode = title]{Toward Open ISA Supercomputing: A Comparative Study of RISC-V, Arm, and x86 Architectures on HPC Workloads
}                      
\tnotemark[1]

\tnotetext[1]{This document is the results of the research
   project funded by the DARE (g.a. 101143421) project, the Italian Research Center on High Performance Computing, Big Data, and Quantum Computing, as well as the FUTURE research project. E.B. and N.S. acknowledge support from the SPACE project, funded by the European Union (g.a. 101093441).}

%\tnotetext[2]{The second title footnote which is a longer text matter
%   to fill through the whole text width and overflow into
%   another line in the footnotes area of the first page.}

\author[1]{Emanuele Venieri}[orcid=0009-0007-7182-505X]

\cormark[1]
\ead{emanuele.venieri2@unibo.it}

%\ead[url]{www.cvr.cc, cvr@sayahna.org}
%\credit{Conceptualization of this study, Methodology, Software}

\author[1]{Giacomo Madella}

\author[1]{Gabriele Ceccolini}
    
\author[1]{Simone Manoni}

\author[2]{Federico Ficarelli}

\author[2]{Filippo Barbari}

\author[3]{Jenny Lynn Almerol}

\author[4]{Elisabetta Boella}

\author[2]{Nitin Shukla}

\author[2]{Simone Bnà}

\author[1,5]{Luca Benini}

\author[1]{Andrea Bartolini}

\affiliation[1]{organization={University of Bologna},
    %addressline={Radarweg 29}, 
    city={Bologna},
    % citysep={}, % Uncomment if no comma needed between city and postcode
    %postcode={1043 NX}, 
    country={Italy}}

\affiliation[2]{organization={CINECA},
    city={Casalecchio di Reno},
    % citysep={}, % Uncomment if no comma needed between city and postcode
    %postcode={695014}, 
    %state={Trivandrum},
    country={Italy}}
\affiliation[3]{organization={University of Trento},
    city={Trento},
    % citysep={}, % Uncomment if no comma needed between city and postcode
    %postcode={695014}, 
    %state={Trivandrum},
    country={Italy}}
\affiliation[4]{organization={E4 Computer Engineering},
    city={Scandiano},
    % citysep={}, % Uncomment if no comma needed between city and postcode
    %postcode={695014}, 
    %state={Trivandrum},
    country={Italy}}   
\affiliation[5]{organization={ETH},
    city={Zurich},
    % citysep={}, % Uncomment if no comma needed between city and postcode
    %postcode={695014}, 
    %state={Trivandrum},
    country={Switzerland}}

\cortext[cor1]{Corresponding author}

%\fntext[fn1]{This is the first author footnote. but is common to third
%  author as well.}
%\fntext[fn2]{Another author footnote, this is a very long footnote and
%  it should be a really long footnote. But this footnote is not yet
%  sufficiently long enough to make two lines of footnote text.}

%\nonumnote{This note has no numbers. In this work we demonstrate $a_b$
%  the formation Y\_1 of a new type of polariton on the interface
%  between a cuprous oxide slab and a polystyrene micro-sphere placed
%  on the slab.
%  }

\begin{abstract}
RISC-V is emerging as a candidate architecture for high-performance computing (HPC), but its competitiveness against established x86 and Arm processors remains difficult to assess because existing studies use different platforms, workloads, and methodologies, and rarely consider energy efficiency. Our work provides an evaluation of contemporary HPC RISC-V, Arm, and x86 processors using a common software stack and methodology. We characterize the Sophgo SG2044 and compare it with the NVIDIA Grace CPU Superchip and Intel Xeon Platinum 8480+ across four HPC benchmarks (STREAM, HPL, HPL-MxP, and HPCG) and four representative scientific and industrial applications (iPIC3D, PLUTO, OpenFOAM, and llama.cpp). The version of HPL-MxP linked to OpenBLAS built and used for this work is made available as open-source software.
The SG2044 reaches 8.02~GFLOP/s of FP64 performance and 7.8~GB/s of memory bandwidth per core, increasing to 294~GFLOP/s and 107~GB/s at 64 cores. L2-cache-aware process placement enables near-linear scaling up to 16 cores, beyond which memory-bandwidth saturation limits further gains. Across generations of RISC-V hardware, in the context of the Monte Cimone (MC) project, the SG2044-based MCv3 doubles both single-core performance and peak memory bandwidth relative to MCv2, while improving memory bandwidth, node-level FP64 performance, and energy efficiency by approximately $100\times$, $158\times$, and $10\times$, respectively, over MCv1.
Against the Arm and x86 systems, the SG2044 shows 8-15$\times$ less absolute performance, reflecting its narrower vector resources and less advanced technology node, but delivers competitive scaling and energy efficiency on several workloads. Its most notable result is LLM decode, where 64-core execution provides the highest relative scaling and energy efficiency among the evaluated platforms while exceeding Intel in throughput. Our results show that current RISC-V processors are closing the gap with the competition among diverse HPC workloads.
\end{abstract}

% \includegraphics{figs/grabs.pdf}

\begin{comment}
\begin{highlights}
\item Extended characterization of the Sophgo SG2044 RISC-V processor across HPC benchmarks, scientific and industrial applications.
\item SG2044 reaches 294 GFLOP/s FP64 and 107 GB/s memory bandwidth at 64 cores, with L2-aware pinning improving scalability.
\item SG2044 matches or exceeds Sapphire Rapids energy efficiency across several workloads and outperforms NVIDIA Grace in LLM inference at 64 cores, reaching 1.20 tok/J (PP) and 0.209 tok/J (TG)
\item Monte Cimone evolution in four years demonstrate up to $158\times$ higher FP64 performance and $10\times$ higher energy efficiency from MCv1 to MCv3.
\item Performance analysis identifies vector width, memory bandwidth, and software-level affinity as key factors for improving RISC-V competitiveness in HPC.
\end{highlights}
\end{comment}

% Keywords

\maketitle
\input{intro}

\input{relwork}
\input{background}

\input{methodology}
\input{results}

\input{conclusions}

%\appendix
%\section{My Appendix}
%Appendix sections are coded under \verb+\appendix+.

%\verb+\printcredits+ command is used after appendix sections to list 
%author credit taxonomy contribution roles tagged using \verb+\credit+ 
%in frontmatter.


\bibliographystyle{unsrt}

\bibliography{cas-refs}

\begin{comment}

\bio{}
Author biography without author photo.
Author biography. Author biography. Author biography.
Author biography. Author biography. Author biography.
Author biography. Author biography. Author biography.
Author biography. Author biography. Author biography.
Author biography. Author biography. Author biography.
Author biography. Author biography. Author biography.
Author biography. Author biography. Author biography.
Author biography. Author biography. Author biography.
Author biography. Author biography. Author biography.
\endbio

\bio{figs/pic1}
Author biography with author photo.
Author biography. Author biography. Author biography.
Author biography. Author biography. Author biography.
Author biography. Author biography. Author biography.
Author biography. Author biography. Author biography.
Author biography. Author biography. Author biography.
Author biography. Author biography. Author biography.
Author biography. Author biography. Author biography.
Author biography. Author biography. Author biography.
Author biography. Author biography. Author biography.
\endbio

\bio{figs/pic1}
Author biography with author photo.
Author biography. Author biography. Author biography.
Author biography. Author biography. Author biography.
Author biography. Author biography. Author biography.
Author biography. Author biography. Author biography.
\endbio
\end{comment}
\end{document}

%% file: intro.tex
\section{Introduction}
The open-source instruction set architecture (ISA) RISC-V has seen remarkable growth in recent years, evolving from a research-oriented initiative into a promising alternative to established proprietary ISAs \cite{venieri2025monte, Brown2025_SG2044}. Its open and modular specification has attracted contributors from academia, startups, and established semiconductor vendors, accelerating the pace of hardware innovation and software ecosystem maturation \cite{venieri2025monte}. As implementations have become more capable and toolchain support has improved, RISC-V platforms have progressed well beyond microcontroller-class devices, giving rise to Linux-capable processors, multi-core system-on-chip (SoC), and accelerators targeting vector processing~\cite{perotti_ara,vitru}, artificial intelligence (AI)~\cite{cavagna_tenstorrent}, and scientific computing workloads. Compiler and runtime support have advanced substantially, with mainstream toolchains such as GCC and LLVM providing full RISC-V backends and vectorization support, while HPC libraries and programming models such as OpenMP, MPI and BLAS are increasingly being ported and optimized for the architecture \cite{Brown2025_SG2044, venieri2025monte}.
This convergence of open hardware and software ecosystems made it feasible to consider RISC-V as an emerging target platform within the context of high-performance computing (HPC) \cite{Brown2025_SG2044, earth_long_vectors, venieri2025monte, almerol2026riscvreadymassivelyparallel, SCW24}, a field traditionally dominated by Intel’s x86 and, more recently, Arm architectures \cite{MANTOVANI2020800}.

In recent years, many research efforts have evaluated Arm and x86 systems for parallel and scientific computing \cite{MANTOVANI2020800, cross-mantovani, ofoam24, AREWEREADY4ARM, PASC19, FIRST_GRACE, uarch_comp, Li2026SPECCC}, and more recently RISC-V to a limited extent \cite{Brown2025_SG2044, venieri2025monte, bartolini2022mcv1, GOMEZ2023152, LAZZERI2026103871, ofoam24}. Nevertheless, the existing body of work remains fragmented, as prior studies generally evaluate only one or two processor families and often focus on limited benchmark suites, individual applications, or specific architectural aspects, making direct cross-architecture comparisons difficult. 
More critically, while power consumption and thermal envelopes represent primary constraints in modern supercomputing, existing comparative evaluations largely overlook comparative energy and power-efficiency analyses, often omitting metrics such as energy-to-solution or performance-per-watt across real-world codes.
Consequently, a unified characterization of contemporary HPC-oriented processors spanning x86, Arm, and RISC-V under a common evaluation methodology remains largely unexplored, preventing a clear understanding of where HPC RISC-V currently stands compared with mature commercial architectures.

To enable a fair and meaningful comparison, this work adopts a consistent experimental methodology across all evaluated platforms. We employ the same compiler toolchain, software libraries, and programming environments whenever possible, while enabling architecture-specific optimizations only when they are mature, publicly available, and representative of real-world usage. The evaluation covers a broad spectrum of HPC-relevant workloads, including synthetic benchmarks, and representative scientific and AI applications. %Specifically, we evaluate \textbf{STREAM} \cite{stream} to characterize sustainable memory bandwidth and the efficiency of memory subsystems, \textbf{HPL} \cite{hpl} and \textbf{HPL-MxP} \cite{hpl-mxp} to assess dense linear algebra performance and the impact of modern low-precision computing capabilities, and \textbf{HPCG} \cite{dongarra2015hpcg} to capture performance in irregular sparse computations and memory-bound workloads representative of many scientific applications. Beyond benchmark suites, we consider real-world applications including \textbf{iPIC3D} \cite{iPIC3D}, a particle-in-cell plasma simulation code representative of large-scale physics workloads; \textbf{PLUTO} \cite{PLUTO}, a computational astrophysics framework for magnetohydrodynamics and high-performance numerical simulations; \textbf {OpenFOAM} \cite{openfoam_original} \cite{openfoam}, a widely used computational fluid dynamics application with complex memory and communication patterns; and \textbf{Llama.cpp} \cite{llama_cpp}, representing emerging AI inference workloads on general-purpose processors.

Specifically, we evaluate STREAM \cite{stream}, HPL \cite{hpl}, HPL-MxP \cite{hpl-mxp}, HPCG \cite{dongarra2015hpcg}, iPIC3D \cite{iPIC3D}, PLUTO \cite{PLUTO}, OpenFOAM, \cite{openfoam_original} \cite{openfoam} and Llama.cpp \cite{llama_cpp}. These workloads cover a broad range of computational characteristics, allowing us to characterize memory bandwidth, dense and sparse linear algebra, low-precision computing, and real-world scientific simulations in plasma physics, astrophysics, and computational fluid dynamics, as well as emerging AI inference workloads on general-purpose processors.
Together, these workloads provide a broad view of computational throughput, scalability, memory behavior, and energy efficiency across different architectures.

In summary, this paper presents an extensive experimental study of contemporary RISC-V processors and their position within the HPC ecosystem. The main contributions of this work are as follows:
\begin{itemize}
\item We provide an extended characterization of the most powerful general purpose, HPC-grade RISC-V CPU chip currently available in the market: the Sophgo SG2044.
%Sophgo SG2044 RISC-V processor, covering
We covered its architectural features, memory subsystem, scalability across 64 cores, and energy efficiency under HPC-relevant workloads. The SG2044 achieves 7.8~GB/s of single-core memory bandwidth and 8.02~GFLOP/s of FP64 performance, reaching 107~GB/s and 294~GFLOP/s, respectively, at 64 cores. L2-cache-aware process pinning improves scaling, reaching twice the bandwidth at 16 cores and revealing memory-bandwidth saturation beyond that core count.

\item We analyze the evolution of RISC-V HPC platforms within the Monte Cimone (MC) research project \cite{bartolini2022mcv1,venieri2025monte}, providing a cross-generational perspective on architectural improvements and their impact on HPC workloads. Compared with MCv2, the SG2044-based MCv3 doubles both the available peak memory bandwidth and single-core performance. Relative to MCv1, memory bandwidth, per-node FP64 performance, and energy efficiency improve by approximately $100\times$, $158\times$, and $10\times$, respectively.

\item We perform a comparison of the SG2044 with HPC-class processors, namely the Arm-based NVIDIA Grace CPU Superchip and the x86-based Intel Xeon Platinum 8480+. While the SG2044 remains 8-15$\times$ behind in absolute performance for several workloads, it achieves competitive or superior scaling in multiple cases while maintaining comparable energy efficiency. In particular, for LLM inference decode, the SG2044 at 64 cores achieves the highest relative scaling ($22.2\times$ vs. $15.7\times$ and $6.1\times$) and energy efficiency among the evaluated platforms (0.209 tok/J vs. 0.159 tok/J and 0.017 tok/J), while also providing higher throughput than the Intel platform (17.4 tok/s vs. 14.0 tok/s).

\item We identify architectural characteristics that influence the competitiveness of current RISC-V processors for HPC, highlighting both their strengths and remaining limitations. For compute-bound workloads such as HPL, the normalized single-core performance of the evaluated platforms follows a trend consistent with the relative width of their vector units, suggesting that increased vector width represents one potential direction for improving single-core performance in future RISC-V HPC processors.

\item HPL-MxP OpenBLAS implementation: we provide an open-source, ready-to-use implementation of HPL-MxP based on the official reference code and linked against OpenBLAS, offering a simple and portable solution that can be readily deployed on generic CPU-based systems, without relying on platform-specific or vendor-optimized implementations.
\end{itemize}
The remainder of this paper is organized as follows. Section \ref{sec:rel_work} describes the context of our evaluation. Section \ref{sec:systems} describes the hardware platforms and their relevant architectural characteristics. Section \ref{sec:software} presents the evaluated benchmarks and applications, together with the experimental methodology adopted for the analysis. Section \ref{sec:results} reports and discusses the experimental results, including performance, scalability, and energy-efficiency comparisons across the evaluated architectures. Finally, Section \ref{sec:conclusions} summarizes the main findings and discusses future research directions.

%% file: relwork.tex
\section{Related Works} \label{sec:rel_work}
In recent years, several studies \cite{PASC19,AREWEREADY4ARM,FIRST_GRACE,uarch_comp,cross-mantovani} have investigated the evolution of HPC processors beyond the traditional x86 ecosystem, with particular attention to Arm-based alternatives. 
\subsection{Arm and x86 HPC Characterization}
\textit{Jackson et al.} \cite{PASC19} provided one of the first evaluations of Arm for HPC, 
%analyzing the Marvell ThunderX2 platform using scientific applications and multi-node workloads. Their study showed that ARM systems could achieve competitive performance with established x86 platforms, with several applications reaching comparable performance while highlighting the importance of software ecosystem maturity.
showing that Arm systems could achieve competitive performance with established x86 platforms and proving its maturity.

More recent works evaluated newer generations of Arm processors, including cloud-oriented systems and HPC-focused designs. \textit{Simakov et al.} \cite{AREWEREADY4ARM} compared ARM systems such as AWS Graviton2/3, Fujitsu A64FX, Ampere Altra, and ThunderX2 against Intel and AMD x86 platforms using HPC benchmarks and scientific applications. 
%Their results showed that modern ARM platforms can achieve competitive performance across a broad range of HPC workloads while frequently providing higher energy efficiency than comparable x86 systems. For example, the study reported that ARM systems achieved comparable performance to x86 platforms in applications such as OpenFOAM, GROMACS, and NWChem, while providing improved energy efficiency in most evaluated cases. However, the performance gap varied significantly across workloads: memory-bandwidth-bound applications benefited from the high core counts and efficient memory subsystems of ARM platforms, whereas workloads relying heavily on optimized x86 libraries or peak floating-point throughput still favored established x86 processors.
They found that modern Arm systems can achieve competitive performance and often higher energy efficiency than x86, although the advantage depends strongly on workload characteristics: memory-bandwidth-bound applications benefited from Arm's high core counts and efficient memory subsystems, while workloads relying on optimized x86 libraries or higher floating-point throughput favored x86 processors.

The recent introduction of NVIDIA Grace CPU Superchip has further increased interest in Arm for HPC-class systems. \textit{Simakov et al.} \cite{FIRST_GRACE} evaluated Grace using HPCC, HPCG, and scientific applications including GROMACS, OpenFOAM, and ROMS, comparing it with Intel and AMD systems. They showed that Grace can achieve competitive node-level performance thanks to its high core count and memory subsystem, matching or exceeding selected x86 systems in several workloads while remaining behind in others. 
%(e.g., matrix multiplication performance was approximately 17\% lower than Intel Sapphire Rapids with HBM). 
From a microarchitectural perspective, \textit{Laukemann et al.} \cite{uarch_comp} compared Arm Neoverse V2, Intel Sapphire Rapids, and AMD Genoa through in-core modeling and microbenchmarks, analyzing execution throughput, memory behavior, and architectural features affecting performance. Similarly, \textit{Banchelli et al.} \cite{cross-mantovani} investigated power efficiency differences between Grace and Sapphire Rapids using STREAM-like and floating-point microbenchmarks, reporting up to 3.25$\times$ higher power efficiency for Grace in selected kernels.

% However, these evaluations have primarily focused on the comparison between two established processor ecosystems, namely x86 and ARM. Over the years, ARM has progressively established itself as a viable alternative to x86 in HPC, demonstrating competitive performance and energy efficiency across a broad range of scientific workloads. The emergence of RISC-V introduces a new dimension to this processor landscape. Its open ISA model, extensible architecture, and the growing involvement of both research and industrial communities have positioned RISC-V as a promising candidate for future HPC systems.
While these evaluations primarily compare the established x86 and Arm ecosystems, the emergence of RISC-V introduces a new dimension. Its open ISA, extensible architecture, and growing community support position RISC-V as a promising candidate for future HPC systems.

\subsection{RISC-V Evaluation for HPC}
Compared with ARM and x86, RISC-V represents a relatively young ecosystem in HPC. Although several high-performance RISC-V platforms have been announced by major vendors, the Sophgo SG2042 and its successor, the SG2044, are currently the only commercially available processors targeting this market. Recent studies \cite{GOMEZ2023152,LAZZERI2026103871,ofoam24,MARQUESGARCIA2026108242,Brown2025_SG2044,BrownJamieson2024_SG2042,almerol2026riscvreadymassivelyparallel} have investigated their potential from different perspectives, including synthetic benchmarks and scientific applications.%, and complete HPC system deployments.

A first class of works focuses on synthetic benchmarks and architectural characterization. \textit{Gomez et al.} \cite{GOMEZ2023152} investigated long-vector architectures for HPCG, including RISC-V vector extensions, proposing optimized implementations of the benchmark kernels. Their evaluation demonstrated that RISC-V vector architectures can effectively exploit memory bandwidth and achieve promising performance for HPCG-like workloads. However, the evaluated RISC-V platform was the European Processor Initiative (EPI) EPAC experimental accelerator rather than a commercially available processor, limiting the direct applicability of the results to production HPC systems. Similarly, \textit{Lazzeri et al.} \cite{LAZZERI2026103871} evaluated several RISC-V processors using CoreMark, Geekbench~6, and SPEC CPU2017, highlighting the progress of recent commercial implementations while showing that a substantial performance gap still exists with respect to contemporary high-end Arm and x86 processors. Collectively, these studies provide valuable insights into specific architectural characteristics, such as vectorization capabilities, memory behavior, and processor efficiency, but rely primarily on synthetic benchmark suites rather than representative HPC workloads.

Complementary efforts have instead considered application oriented evaluations. \textit{Suarez et al.} \cite{ofoam24} compared Arm and RISC-V SoCs using performance and energy measurements, employing OpenFOAM as a representative computational fluid dynamics application. While the study highlighted the energy-efficiency potential of RISC-V implementations, the evaluated systems targeted embedded-class platforms rather than HPC-oriented processors. More recently, \textit{Garcia et al.} \cite{MARQUESGARCIA2026108242} investigated RISC-V vector extensions for large language model inference on the commercially available Sophgo SG2042 CPU, showing that RVV can accelerate inference by up to 1.3$\times$ depending on workload characteristics while also exposing current software ecosystem limitations. Although these works demonstrate the applicability of RISC-V to real applications, they remain focused on individual application domains and do not provide a broader characterization across the diverse workload spectrum encountered in HPC.

Finally, recent efforts have started evaluating complete server-class RISC-V platforms. \textit{Brown} \cite{Brown2025_SG2044} presented the first characterization of the Sophgo SG2044 processor through STREAM and NAS Parallel Benchmarks, comparing it with AMD EPYC, Intel Skylake, and Arm ThunderX2 systems. The study highlighted the substantial improvements of the SG2044 memory subsystem and its scalability compared with the previous SG2042 generation. However, its evaluation primarily focused on micro-benchmarks and pseudo-applications, without an in-depth analysis of the SG2044's UMA architecture or its behavior across a broader set of real-world HPC applications. 
\subsubsection{The Monte Cimone Project}
\begin{comment}
Complementing these efforts, the Monte Cimone project \cite{bartolini2022mcv1, venieri2025monte} has progressively explored the adoption of RISC-V for HPC through successive hardware generations, evolving from an initial hardware--software co-design platform into a research infrastructure for evaluating increasingly capable server-class RISC-V processors. Across its different generations, the project has addressed software ecosystem maturity, system integration, compiler support, and performance characterization, providing a longitudinal perspective on the evolution of RISC-V for HPC.
\end{comment}

Complementing these efforts, the Monte Cimone project \cite{bartolini2022mcv1, venieri2025monte} hosted at University of Bologna’s DEI department in collaboration with CINECA and E4 Computer Engineering, has progressively explored the adoption of RISC-V for HPC through successive generations of clustered hardware. Initially conceived as a hardware--software co-design platform, Monte Cimone has evolved into a research infrastructure for evaluating increasingly capable server-class RISC-V processors and their supporting software ecosystem. The platform integrates HPC-oriented components, including the SLURM job scheduler, EXAMON system monitoring, shared home directories, and environment modules for consistent software deployment across nodes. This infrastructure has supported research spanning RISC-V software porting and optimization \cite{venieri2025monte,mittone_mc,nassyr2026,seibel2026}. The successive generations therefore provide a longitudinal view of RISC-V evolution in HPC while serving as a reusable platform for developing and evaluating the associated software stack.

Overall, existing studies have significantly advanced the understanding of RISC-V for HPC from architectural, application, and system-level perspectives. Nevertheless, the current literature remains fragmented across different processor generations, benchmark suites, and application domains. As a result, it is still difficult to assess the current position of RISC-V within the HPC processor landscape, particularly with respect to contemporary x86 and Arm server-class processors. A cross-ISA characterization based on modern representatives of the three ecosystems, evaluated under a common experimental methodology and across a broad spectrum of synthetic, scientific, and AI workloads, is still missing.

%% file: background.tex
\section{Systems and Platforms} \label{sec:systems}
\begin{figure}[]
    \includegraphics[width=\columnwidth]{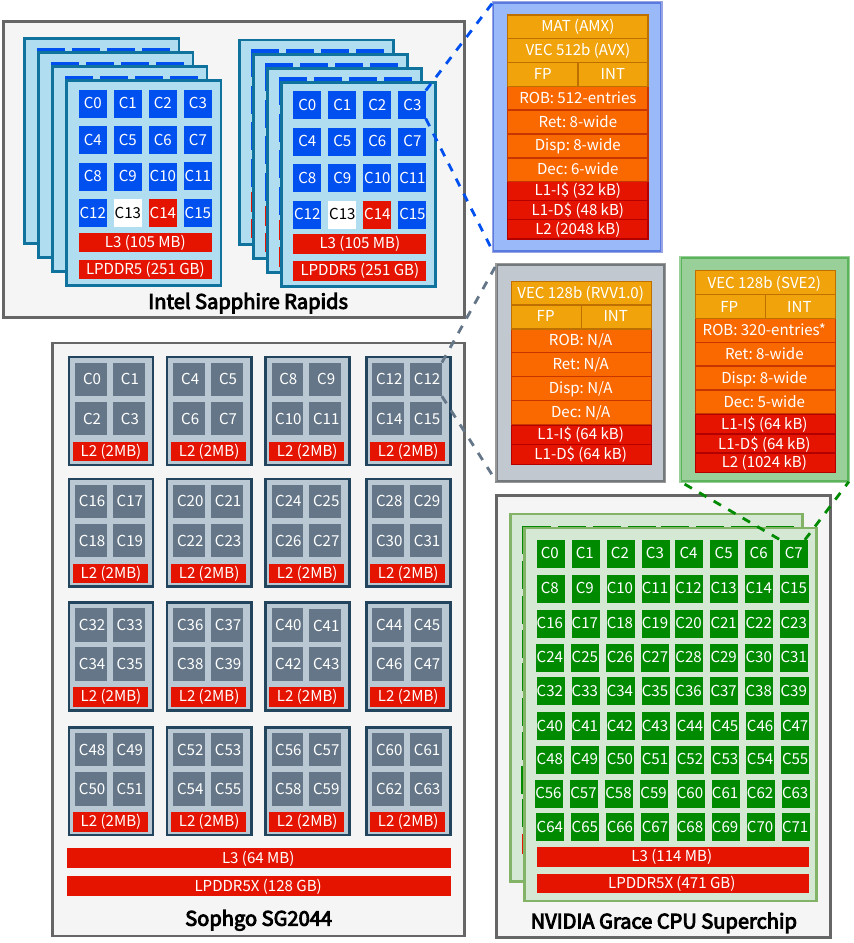}
    \caption{Overview of the three server-class processors evaluated in this work. The figure compares the Intel Xeon Platinum 8480+ (Sapphire Rapids), NVIDIA Grace CPU Superchip, and Sophgo SG2044 architectures, highlighting their core organization, cache hierarchy, vector execution capabilities, memory subsystem, and node-level configuration.}
    \label{fig:architecture_overview}
\end{figure}

\begin{table}[]
\centering
\scriptsize
\caption{Hardware characteristics of the evaluated platforms.}
\begin{tabular}{lccc}
\hline
 & SG2044 & Intel SR & NVIDIA GS \\
\hline
ISA & RISC-V & x86-64 & Armv9 \\
Cores / Node & 64 & 112 (2×56) & 144 (2×72) \\
Vector ISA & RVV 1.0 & AVX-512 & SVE2 \\
Vector length & 128 bits & 2$\times$512 bits & 4$\times$128 bits \\
Technology node & TSMC 12nm & Intel 7 & TSMC 4N \\
Max Frequency [GHz] & 2.6 & 3.8 & 3.3\\
Memory channels & 32 & 16 & 32\\
Memory Type & LPDDR5X & DDR5 & LPDDR5X \\
Memory & 128 GB & 2 TB & 960 GB \\
NUMA & No & Yes & Yes \\
\hline
\end{tabular}
\label{tab:systems}
\end{table}
To assess the current state of RISC-V for HPC, we compare representative server-class processors from the three SoA major CPU ecosystems: x86, Arm, and RISC-V. The evaluated systems are all commercially available platforms, avoiding comparisons against experimental prototypes or pre-production hardware. For the proprietary ISAs, we select Intel Sapphire Rapids and NVIDIA Grace, as they represent two of the most relevant server processors currently deployed in HPC systems. Sapphire Rapids is the most widely deployed Intel processor generation in the June 2026 TOP500 list, powering 84 systems (16.8\% of the ranked machines), while NVIDIA Grace is the most widely deployed Arm processor in the same list, appearing in 25 systems (5\%) \cite{TOP5002026}. Grace also represents NVIDIA's latest Arm-based CPU architecture specifically designed for HPC and AI workloads. For RISC-V, we evaluate the Sophgo SG2044, which is currently the most advanced commercially available high-core-count server processor based on the RISC-V ISA \cite{Brown2025_SG2044, venieri2025monte}. 

This selection enables a meaningful assessment of the current position of RISC-V with respect to the highest-performance commercially available processors from the established x86 and Arm ecosystems.

\subsection{Architectural Overview}
The main architectural characteristics of the evaluated platforms are summarized in Fig.~\ref{fig:architecture_overview} and described below.

\subsubsection{Intel Xeon Platinum 8480+} (\emph{Intel SR}): a Sapphire Rapids generation x86\_64 processor. Official Intel documentation describes the ISA features and cache hierarchy, whereas front-end characteristics (6-wide decode, 8-wide dispatch and retirement, and a 512-entry reorder buffer) follow publicly available microarchitectural analyses \cite{intel_opt_manual,golden_cove}. Each Golden Cove core integrates AVX-512 vector extensions and Intel Advanced Matrix Extensions (AMX), together with private 48~KB L1 data and 2~MB L2 caches. The evaluated system consists of two sockets which each of them comprises four Compute Tiles interconnected through Intel's on-chip mesh, exposing four NUMA domains and sharing a 105~MB L3 cache. Each tile physically integrates 16 cores, of which one serves as a memory core, while one additional core per socket is inactive due to silicon manufacturing defects (highlighted in red and white, respectively, in Fig.~\ref{fig:architecture_overview}), resulting in 56 usable cores per socket (112 in total). Hyper-Threading was disabled throughout the experiments and frequency scaling governor set to \textit{performance} to ensure a consistent comparison across architectures and to focus the analysis on physical-core performance.

\subsubsection{NVIDIA Grace CPU Superchip} (\emph{NVIDIA GS}): a dual-socket Armv9 platform comprising two 72-core processors, for a total of 144 Neoverse V2 cores. Official Arm and NVIDIA documentation describe the processor architecture, cache hierarchy, and SVE2 ISA support, whereas front-end characteristics (6-wide decode, 8-wide dispatch, 8-wide retirement, and a 320-entry reorder buffer) follow the values reported in \cite{arm_hotchips23}. Each core provides private 64~KB L1 data and 1~MB L2 caches, while a 114~MB L3 cache is shared by the 72 cores within each die. The two Grace dies form two NUMA domains connected through the NVLink-C2C interconnect, resulting in non-uniform memory access between the two sockets.

\subsubsection{Sophgo SG2044} (\emph{SG2044}): a 64-core RISC-V processor implementing the RV64GCV ISA with RVV 1.0 vector support. The processor is composed of 16 four C920V3 core clusters. The C920V3 core features a 12-stage out-of-order pipeline with a 4-wide issue width and a 10-wide execution architecture, as reported by the IP manufacturer \cite{xuantie_c920}. Each core provides a private 64~KB L1 data cache, while groups of four cores share a 2~MB L2 cache and all cores share a 64~MB L3 cache. Unlike the x86 and Arm platforms, the SG2044 exposes a Uniform Memory Access (UMA) architecture, in which all cores observe approximately the same latency when accessing any region of the 128~GB LPDDR5X main memory. The processor is deployed within the Sophgo RISC-V Compute Server SRA3-40 platform \cite{sophgo-server}.
The main characteristics of each platform are summarized in Table \ref{tab:systems}.

%% file: methodology.tex
\section{Evaluation Methodology} \label{sec:software}

The methodology adopted in this work is designed to minimize software-induced variability across the evaluated platforms, enabling a more direct assessment of performance differences attributable to architectural and system-level characteristics. To this end, we rely on widely available, community-supported software stacks and apply the same compiler toolchains, libraries, and runtime configurations whenever possible. Vendor-specific benchmark suites or proprietary optimizations that could introduce unfair advantages are avoided; however, mature and publicly available optimizations required to fully exploit each architecture (e.g., vector instruction support and optimized numerical libraries) are enabled when representative of HPC practice.

To establish a consistent experimental baseline, all systems run the Linux kernel 6.12 and use GCC 14.2 as the reference compiler. The Message Passing Interface (MPI) layer is provided by OpenMPI 5.0 across all platforms. As a reference high-performance linear algebra library, OpenBLAS 0.3.34 is employed, compiled with architecture-specific optimization flags as reported in table \ref{tab:openblas_config}. This configuration is used for all relevant benchmarks, including dense linear algebra workloads such as HPL.

\begin{table}[]
\footnotesize
\centering
\caption{OpenBLAS compilation target parameter for each machine}
\begin{tabular}{ll}
\hline
\textbf{Machine} & \textbf{TARGET parameter}\\
\hline
SG2044 & RISCV64\_ZVL128B \\
Intel SR & SAPPHIRERAPIDS \\
NVIDIA GS & NEOVERSEV2 \\
\hline
\end{tabular}
\label{tab:openblas_config}
\end{table}

% Power consumption is measured mainly using the telemetry exposed by the Baseboard Management Controller (BMC) through IPMI on all evaluated systems.  
% The Intel Sapphire Rapids platform additionally provides processor-level power monitoring through Intel Running Average Power Limit (RAPL), while NVIDIA Grace exposes platform-specific power telemetry through NVIDIA management and hardware monitoring (hwmon) interfaces . However, these architecture-specific mechanisms expose different power domains and are not consistently available across all platforms. In particular, the SG2044 does not provide an equivalent processor-level power monitoring interface and relies on BMC/IPMI telemetry for system-level measurements. Therefore, to ensure a consistent comparison across x86, Arm, and RISC-V platforms, we adopt BMC/IPMI measurements as the common power monitoring methodology for all evaluated systems.
To obtain power measurements, we first collected all available sensor data from the evaluated systems through the ExaMon \cite{examon_bartolini} infrastructure. On x86 and Arm nodes, IPMI telemetry was used to measure total node power consumption and fan speed. 
In addition, architecture-specific mechanisms were exploited where available: Running Average Power Limit (RAPL) on the Intel Sapphire Rapids and hardware monitoring (hwmon) on NVIDIA Grace were used to collect processor-level power measurements. 
A key difference concerns the RISC-V SG2044, IPMI reports only board-level power and excludes fans and other auxiliary components.
We therefore engineered combined power metrics by modeling the power consumption of fans and auxiliary components for each architecture and combining these models with RAPL and hwmon measurements, allowing us to isolate board-level power and define a common power domain across all platforms. 

\subsection{NUMA aware process pinning}\label{sec:pinnings}
For each benchmark, we ran experiments on process-to-core pinning, aiming to find an optimal placement that reflects the machine's memory topology and to characterize the
differences with respect to other strategies.
In Table \ref{tab:placement_policies} we summarize the adopted pinning strategies.

\begin{table*}[]
    \footnotesize
    \centering
    \caption{Thread/process placement policies.}
    \label{tab:placement_policies}
    \begin{tabular}{llp{0.60\textwidth}}
        \hline
        \textbf{Platform} & \textbf{Policy} & \textbf{Description} \\
        \hline
        All & \texttt{os\_default} &
        OS scheduler free to migrate processes. \\
        
         & \texttt{sequential} &
        processes pinned in raw core order (0, 1, 2, ...) \\
        \hline
        
        SG2044 & \texttt{l2\_cache\_aware} &
        (best) round-robin across all 16 L2 clusters, stride 4 (0, 4, 8, 12, ...). \\
        \hline
        
        Grace & \texttt{dual\_die\_interleaved} &
        (best) round-robin between the two dies (0, 72, 1, 73, ...).
        caps total bandwidth to one die. \\
        \hline
        
        Sapphire Rapids & \texttt{numa\_aware\_distributed} &
        (best) balancing processes across NUMA nodes on both sockets. \\
        
        & \texttt{numa\_aware\_local} &
        fills socket 0's physical cores first, then socket 1 (0, 2, 4, ..., 110, 1, 3, ..., 111) \\
        
        & \texttt{strict\_numa\_local} &
        fills one NUMA node completely before the next, socket 0 first then socket 2. \\
        
         & \texttt{strict\_numa\_distributed} &
        fills one NUMA node of socket 0, then one of socket 1, alternating. \\
        \hline
    \end{tabular}
\end{table*}

On SG2044, which is a UMA machine, the \texttt{l2\_cache\_aware} strategy generally proved to be the best. By distributing processes evenly across the 16 L2 clusters, saturation of the L2 caches is 
delayed, allowing the workload to benefit longer from the higher aggregate L2 bandwidth. Once saturation is reached, the advantage shrinks and becomes equivalent to the \texttt{sequential} strategy.

On Grace, the best strategy is to alternate process pinning across both NUMA nodes (the two dies), so that all of the chip's memory controllers are engaged from the start, fully exploiting the bandwidth available to the whole system. On Sapphire Rapids, similarly to Grace, the \texttt{numa\_aware\_distributed} strategy proved to be the best. 

%It is also worth noting that in some cases, for example OpenFOAM on Intel Sapphire Rapids showed in Figure \ref{fig:openfoam_pinning_efficiency}, the performance obtained in the absence of pinning, that is, with free process allocation and migration left to the OS, comes very close to the performance of the best pinning strategies found experimentally. This suggests that the OS autonomously chooses placement strategies similar to ours. By distributing processes across the NUMA nodes of both sockets in a balanced way from the start, all memory controllers are engaged, fully exploiting the system's available bandwidth.

Overall, we found that on our NUMA platforms it is preferable to use all available bandwidth from the start, even at the cost of placing processes on different NUMA nodes, thus paying the higher latency of inter-process MPI communication across NUMA node boundaries. On the UMA machine, however, it is still worth balancing processes across the L2 clusters, so as to avoid prematurely saturating the L2 caches.

% \subsection{Strong scaling methodology}
\subsection{Benchmark Suite and Applications}
As evaluation software, we combine synthetic benchmarks with real-world applications operating on comparable data formats and computational primitives to provide a balanced characterization of both peak performance and system-level behavior. Synthetic benchmarks isolate specific architectural characteristics, such as floating-point throughput, memory bandwidth, and energy efficiency, while application level workloads provide a representative assessment of end-to-end performance across scientific simulations and AI workloads.
The specific parameters used for each test are described in the corresponding sections. We evaluate strong scaling: for a given problem size, the workload is kept constant while increasing the number of active cores. This allows us to assess how effectively each platform exploits additional compute resources for the same computational problem.

\subsubsection{Synthetic Benchmarks}
The synthetic benchmark suite includes the components explained in the following paragraphs.

\paragraph{\textbf{STREAM:}} Evaluates the sustainable memory bandwidth of a system using four fundamental vector operations (Copy, Scale, Add, and Triad) executed over large, contiguous data arrays \cite{stream}. Unlike compute-bound workloads, STREAM is heavily memory-bound, featuring an arithmetic intensity near zero with $O(N)$ computational complexity scaling directly with $O(N)$ data movement. %The four kernels, mathematically defined as $a_i = b_i$ (Copy), $a_i = q \cdot b_i$ (Scale), $a_i = b_i + c_i$ (Add), and $a_i = b_i + q \cdot c_i$ (Triad), are deliberately designed to minimize arithmetic overhead, ensuring that execution time is strictly dictated by the data transfer rate between main memory and the processor. 
Performance is reported as sustained memory bandwidth in gigabytes per second (GB/s) for each kernel. 

The benchmark is configured according to the general recommendations provided in the official STREAM guidelines. Since our objective is to characterize main-memory performance, we set \texttt{STREAM\_ARRAY\_SIZE} to 120,000,000, resulting in 960 MB per array when using double-precision elements. The selected size ensures that the STREAM working set substantially exceeds at least four times the Last-Level Cache (LLC) capacity of all evaluated systems, thereby minimizing the influence of cache effects on the measured memory bandwidth.
The \texttt{NTIMES} parameter is kept at its default value of 10. Consequently, each STREAM execution performs 10 timed repetitions, with the best result reported by the benchmark. At a higher level, our measurement pipeline executes each experiment five independent times. The final reported value is the average across these five executions, with the corresponding variability reported as the measurement error.
Among the STREAM kernels evaluated, in section \ref{sec:res_stream} we report the Triad kernel as the representative case in this analysis. %Triad is particularly suitable for investigating the scaling of the memory subsystem, as its performance is primarily determined by sustained memory bandwidth and it combines multiple memory streams with a read-modify-write operation. 
The remaining STREAM kernels were also evaluated and showed consistent qualitative trends; therefore, to avoid redundancy, only Triad results will be discussed.

\paragraph{\textbf{HPL:}} The High Performance Linpack benchmark measures sustained floating-point performance by solving a dense linear system $Ax = b$ through LU factorization with partial pivoting ($P\!A = LU$) \cite{hpl}. Computation is performed in 64-bit floating-point (FP64) and is dominated by the $O(N^3)$ matrix-matrix multiplication kernel (DGEMM), making standard HPL a compute-bound workload and a reference benchmark for evaluating sustained peak throughput ($R_{\text{max}}$) expressed as performance in GFLOP/s. %, calculated strictly over the factorization and back-substitution runtime using the standard operation count of $\frac{2}{3}N^3 + 2N^2$ FLOPS. 
In our evaluation, we utilize the reference HPL 2.3 implementation linked against OpenBLAS (v0.3.34) and using a MPI rank per core.
HPL exposes two main classes of configuration parameters. The first defines the problem geometry and parallel decomposition, including the global matrix size ($N$), block size ($N\!B$), and two-dimensional process grid ($P \times Q$). These parameters directly affect load balancing, communication overhead, and cache utilization, and were experimentally tuned on each platform to maximize sustained performance. The second class controls the solving strategy and to ensure a fair comparison across architectures, these algorithmic parameters were kept identical for all experiments, as summarized in Table~\ref{tab:hpl_config}. Following the HPL recommendations, $N$ was selected to use approximately 70\% of available memory and rounded to the nearest multiple of $N\!B$. The block size was empirically selected through full-node sweeps over $N\!B=64$--$512$.
The process grid follows the HPL recommendation of a nearly square grid with $P \leq Q$ and $P \times Q=N_{\mathrm{MPI}}$, and to match $P$ with the number of processors in multi-CPU systems. For NVIDIA Grace, $P=2$ was fixed to match its two interconnected CPUs, with $Q$ adjusted to satisfy $P \times Q=N_{\mathrm{cores}}$. For Intel Sapphire Rapids, both $P=2$, matching the two CPUs, and $P=8$, matching the NUMA domains, were evaluated, with $Q$ adjusted accordingly. Conversely, the RISC-V platform consists of a single-socket processor ($P=1$), for which multiple executions at the maximum core count were performed to empirically determine the process-grid configuration yielding the highest performance. The final machine-dependent parameters are reported in Table~\ref{tab:hpl_config_exp}.
\begin{table}[t]
\footnotesize
\centering
\caption{HPL.dat parameters. The first group was determined with ad hoc experiments.}
\begin{tabular}{lll}
\hline
\textbf{Category} & \textbf{Parameter} & \textbf{Value} \\
\hline
\multirow{4}{*}{Tuned}
& $N$ & function of memory size \\
& $NB$ & experimental \\
& $P$ &  experimental \\
& $Q$ &  experimental \\
\hline
\multirow{13}{*}{Fixed}
& PMAP & Row-major \\
& Threshold & 16.0 \\
& PFACT & Left \\
& NBMIN & 2 \\
& NDIV & 2 \\
& RFACT & Right \\
& BCAST & 1-ring \\
& DEPTH & 0 \\
& SWAP & Mix \\
& Swap threshold & 64 \\
& L1 storage & Transposed \\
& U storage & Transposed \\
& Equilibration & Enabled \\
& Memory alignment & 8 doubles \\
\hline
\end{tabular}
\label{tab:hpl_config}
\end{table}

\begin{table}[t]
\footnotesize
\centering
\caption{HPL.dat machine-tuned parameters.}
\begin{tabular}{lllll}
\hline
\textbf{Machine} & \textbf{N} & \textbf{NB} & \textbf{P} & \textbf{Q}\\
\hline
SG2044 & 105792 & 64 & 4 & 16 \\
Sapphire & 418176 & 384 & 8 & 14 \\
Grace & 289792 &  128 & 2 & 72 \\
\hline
\end{tabular}
\label{tab:hpl_config_exp}
\end{table}

\paragraph{\textbf{HPCG:}} the High Performance Conjugate Gradient benchmark, complements HPL by evaluating sparse iterative computations that are representative of many scientific applications. It solves a synthetic sparse linear system arising from the discretization of a 3D problem using a preconditioned conjugate gradient solver with a geometric multigrid preconditioner \cite{hpcg_technical}. Unlike HPL, HPCG is primarily limited by irregular memory accesses, data movement, and communication, making it representative of memory-bound scientific workloads. Performance is reported as the sustained floating-point rate (GFLOP/s) during the timed execution.
To evaluate scalability under a constant workload, all experiments employ a strong-scaling configuration. We use a global problem size of $384^3$ cells (approximately 56 million unknowns, corresponding to about 40.5~GB of data on a single process), which is evenly partitioned among the available MPI ranks. The local subdomain dimensions are adjusted as the number of ranks increases to keep the global problem approximately constant. This problem size was selected to ensure that, even at the highest core counts, the local working set remains significantly larger than the available cache capacity, so that the benchmark is primarily limited by the main memory subsystem rather than cache bandwidth.

\paragraph{\textbf{HPL-MxP:}} A mixed-precision variant of the HPL benchmark designed to bridge the gap between traditional double-precision scientific workloads and the low-precision tensor capabilities of modern HPC hardware. To reconcile these divergent hardware trends, HPL-MxP combines modern mixed-precision algorithms with maximum hardware efficiency while maintaining direct numerical parity with standard HPL \cite{Dongarra_2025}. 
The algorithm shifts the computationally dominant $O(N^3)$ LU factorization into a hardware-supported reduced-precision format like FP16, BF16, or FP32. The resulting low-precision factors then serve as a preconditioner within an iterative refinement loop, typically using the Generalized Minimal Residual (GMRES) Krylov method, executed in full FP64 precision to recover standard double-precision accuracy ($\frac{\|Ax - b\|}{\|A\|\|x\| + \|b\|} \le O(\varepsilon_{\text{FP64}})$). Effective throughput is reported in sustained FP64-equivalent FLOPS, defined by dividing the theoretical FP64 operation count (as defined in HPL) by the combined execution time of both the factorization and refinement phases.
% \hl{TODO: which format has been chosem? the which problem size has been chosen (see section 6 of HPL-MxP paper)? why? Which precision are we using on which machine?}
% In the implementation that has been used two-precision scheme has chosen. The experiments use different dimension of the matrix ($N$) based on the available amount of memory on each platform, in particular we tried to stay with > 70\% of memory usage for each of them. The original matrix and right-hand side are generated and stored in FP64. They are then converted to FP32, where the LU factorization without pivoting and the triangular solves are performed. The resulting factors and solution are converted back to FP64, and GMRES-based iterative refinement, residual computation, and final validation are performed in FP64. This precision assignment is identical across all evaluated machines; only the architecture-specific OpenBLAS kernels and compiler optimization flags differ. From the reference HPL-AI implementation %\cite{}
% , we replaced the internal BLAS back-end, including SGEMM, STRSM, DTRSM, and DGEMV, with the shared OpenBLAS library on its $0.3.34$ version. 
Due to the lack of a ready-to-use, multi-architecture solution, we created our own implementation \footnote{HPL-MxP repository used: \url{https://gitlab.com/ecs-lab/tools/hpl-mxp-ecslab}} starting from the reference HPL-MxP code \cite{hpl-mxp}. Specifically, we replaced its internal BLAS backend, including SGEMM, STRSM, DTRSM, and DGEMV, with the shared OpenBLAS library. This approach enabled us to leverage OpenMP without further editing the core algorithms, which we used to exploit parallelism, tailoring it on each architecture. For this same reason, to avoid deviating further from the original reference, we decided to step down only to FP32, implementing a dual-precision scheme based solely on FP64 and FP32.
In accordance with the standard methodology for HPL and HPL-MxP benchmarks, which require stressing the system's memory capacity to properly assess peak performance,we scale the problem size (N) to utilize at least 70\% of the available memory on each platform. The computational workflow begins with the matrix and right-hand side generated in FP64, which are then downcast to FP32 for the core factorization phase. Finally, the results are cast back to FP64 for iterative refinement and validation. This mixed-precision scheme is kept strictly identical across all evaluated machines; only the architecture-specific OpenBLAS kernels and compiler optimization flags differ.
% The experiments use different matrix dimensions ($N$), depending on the amount of available memory on each platform; in particular, as suggested in the offical guide, we aim to use more than 70\% of the available memory on each system. The original matrix and right-hand side are generated and stored in FP64. They are then converted to FP32, where the LU factorization without pivoting and the triangular solves are performed. The resulting factors and solution are subsequently converted back to FP64, and GMRES-based iterative refinement, residual computation, and final validation are performed in FP64. This precision assignment is kept identical across all evaluated machines; only the architecture-specific OpenBLAS kernels and compiler optimization flags differ.

\subsubsection{Applications}
The selected applications span a diverse set of computational domains and execution characteristics, enabling the evaluation of the architectures under representative HPC and AI workloads. Together, they cover a broad range of computational patterns, including particle-based simulations, stencil computations, computational fluid dynamics, and transformer-based inference. To ensure a fair comparison, identical input datasets, software versions, and compilation configurations are used across all evaluated platforms whenever possible.

% \begin{itemize}
% \item \textbf{Astrophysical code}: (?) The code primarily involves numerical integration and grid-based computations, with a mix of memory-bound and compute-bound kernels. It is included to represent irregular scientific workloads typical of computational astrophysics.

\paragraph{\textbf{iPIC3D:}} a semi-implicit Particle-in-Cell (PIC) code developed to simulate collisionless plasma dynamics at kinetic scales \cite{iPIC3D}. The code is written in C/C++ and parallelised via MPI. The simulation advances macro-particles in a Lagrangian framework while solving Maxwell's equations and computing plasma moments, including charge density and current, on an Eulerian grid. %The implicit formulation relaxes the stringent stability constraints of explicit PIC methods, enabling larger timesteps and reducing communication frequency in distributed-memory executions. 
The computational workload is dominated by particle pushing together with particle-to-grid and grid-to-particle interpolation. 
% Although particles are processed sequentially, each particle interacts with a small neighborhood of grid cells whose location changes over time, resulting in irregular memory accesses to the grid-resident field and plasma data and limited data reuse \cite{Madduri}. In addition, particle deposition requires frequent scatter updates to shared grid arrays. 
Particles are traversed in memory order, unrelated to their position in the domain, so unless they are sorted by cell, each particle's interaction with a small, time-varying neighborhood of grid cells leads to irregular memory accesses and limited data reuse \cite{Madduri}. We run iPIC3D in its sorted mode, which groups particles by cell so that consecutive particles access neighboring grid data; this was the best-performing configuration in \cite{astro}.
Overall, iPIC3D is largely memory-bound rather than compute-bound.
The selected 2D simulation models the current filamentation instability, a typical plasma instability which leads to the amplification of the magnetic field in the direction out of the plane~\cite{CFI}.
The computational domain is decomposed only along the $x$ and $y$ dimensions. For a given number of MPI ranks $N$, the process grid $\mathrm{XLEN}\times\mathrm{YLEN}$ is chosen as the factorization of $N$ closest to a square ($\mathrm{XLEN}\leq\mathrm{YLEN}$ and $\mathrm{XLEN}\cdot\mathrm{YLEN}=N$), while $\mathrm{ZLEN}=1$. To ensure an even partitioning of the fixed $256\times256$ grid, both $\mathrm{XLEN}$ and $\mathrm{YLEN}$ must divide 256. Consequently, only rank counts satisfying this constraint are considered, resulting in strong-scaling experiments with 1, 2, 4, 8, 16, 32, and 64 MPI ranks (and 128 on the Grace system).

\paragraph{\textbf{PLUTO:}} a C/C++ code for computational plasma astrophysics that solves the magnetohydrodynamics (MHD) equations with a finite-volume Godunov-type scheme on structured grids, computing interface fluxes via local Riemann solvers \cite{PLUTO}.
% a code for computational plasma astrophysics, implemented in C/C++, that solves the equations of magnetohydrodynamics (MHD) on structured grids using a finite-volume formulation \cite{PLUTO}. In this approach, the computational domain is discretized into control volumes (cells), and the conserved physical quantities are updated by computing the numerical fluxes exchanged across neighboring cell interfaces. These fluxes are evaluated using Godunov-type methods, which solve local Riemann problems at cell boundaries to accurately capture discontinuities such as shocks and contact interfaces.
The computational workload is dominated by stencil-based updates and Riemann solver evaluations. These kernels involve intensive double-precision floating-point operations with structured memory access patterns, making the application largely compute-bound and able to benefit from SIMD vectorization. Parallel execution relies on MPI domain decomposition with halo exchanges between neighboring subdomains, whose cost becomes increasingly significant as the number of processes increases.

We configure PLUTO with the Orszag--Tang vortex, a standard ideal-MHD
test problem, discretized on a Cartesian grid of $336^3$ cells.
The $336 = 2^4\cdot3\cdot7$ admits balanced, near-cubic domain decompositions for every
process count used in our sweeps.
Since we are only interested in the per-iteration
computational cost and not in the physical evolution of the vortex, each run is limited to
a fixed number of 10 integration steps.
We perform a strong-scaling study: the global grid is held fixed at $336^3$ cells while the
process count is varied from 1 up to the full core count of each node.% (64 on SG2044, 144 on Grace, 112 on Sapphire Rapids), with the domain decomposed as close to a cube as the process count and the divisors of 336 allow.

\paragraph{\textbf{OpenFOAM:}} a leading open-source Computational Fluid Dynamics (CFD) framework widely used in scientific and engineering simulations.
Parallelism is exploited exclusively at the MPI level, through domain decomposition and distribution of the resulting subdomains across cores, with inter-process communication at subdomain boundaries.
It relies heavily on sparse linear system solvers and discretizations on unstructured grids, placing significant pressure on memory bandwidth and latency while also stressing inter-core communication in parallel executions.

The tests conducted on OpenFOAM (v2512) were performed on a single case study,
the Windsor Squareback geometry \cite{windsor}, which simulates an
idealized vehicle profile in a wind tunnel and is used in the
Automotive CFD Workshop (AutoCFD, Case~1)~\cite{autocfd_workshop}.
For practicality and computational/time-resource reasons, the smallest
available mesh (grid~1) was chosen, consisting of approximately
6.3~million cells. Despite its modest size, this case study is
qualitatively close to an industrial automotive CFD case, making it
significant for this study.
Domain decomposition across MPI processes is handled by the KaHIP
graph-partitioning algorithm~\cite{kahip}.
% which assigns mesh cells to
% cores while balancing computational load and, at the same time,
% minimizing the number of interface faces between subdomains (faces
% shared between cells assigned to different cores). At every
% linear-solver iteration, these interface faces require a data exchange
% via MPI between the processes involved.
The case is run with simpleFoam, the incompressible finite-volume solver of OpenFOAM, which couples pressure and velocity through the SIMPLE algorithm \cite{SIMPLE}. Every configuration executes a fixed budget of 100 iterations, so that the amount of work is identical across architectures and core counts and the measured wall-clock times are directly comparable. 
Turbulence is modeled using the Reynolds-Averaged Navier–Stokes (RANS) approach.
% , a widely used method that captures the mean effect in time of turbulence without resolving its finest scales, keeping the computation affordable.
The dominant cost of each iteration is the solution of the pressure equation, with one unknown per cell (of the order of six million) and characterized by a large sparse linear system, solved with a geometric-algebraic
multigrid method (GAMG) \cite{multigrid_tutorial} using the Gauss-Seidel smoother \cite{saad2003iterative}.
% The smoother is the most computationally expensive kernel within the GAMG method. It sweeps the rows of the matrix in sequence and, for each row, gathers the coefficients of the neighbouring cells through the indirect indices of the sparse storage, accumulates them with multiply–add operations and divides by the diagonal. It is a sparse, irregular and strongly memory-bound kernel with very low arithmetic intensity, whose performance is dictated by the memory performances rather than by peak floating-point throughput, and which in parallel further requires the exchange of interface values between neighbouring subdomains at every sweep.
% In the upstream version of OpenFOAM used in this study, all sparse matrix operations, including those performed by the smoother, are executed using OpenFOAM's native sparse format, known as LDU-COO. While this format is designed for efficient execution on scalar cores, it is inherently hostile to vectorization and SIMD processing due to sparse memory access patterns and potential race conditions during result updates. Consequently, compiler-driven vectorization is restricted to loops over dense arrays, whose computational weight is negligible compared to the sparse matrix operations. As a result, OpenFOAM's execution is highly parallel across multiple cores via MPI, yet strictly scalar at the individual core level.
In the upstream OpenFOAM version used here, all sparse matrix operations, rely on OpenFOAM's native LDU-COO format. Designed for scalar execution, this format is ill-suited to vectorization because of irregular memory access and potential race conditions during result updates. Compiler auto-vectorization is thus limited to loops over dense arrays, whose cost is negligible. OpenFOAM is therefore highly parallel across cores via MPI, but strictly scalar within each core.

\paragraph{\textbf{llama.cpp:}} a C++ implementation of large language model (LLM) inference optimized for CPU execution. It enables efficient execution of transformer-based models in environments without dedicated accelerators. While LLM inference is typically deployed on GPU-based systems in large-scale, high-throughput settings, CPU inference remains relevant for single-user or low-concurrency scenarios, particularly when large models must be executed within the constraints of memory capacity and without multi-GPU model parallelism.
In these conditions, the absence of batching and high request concurrency shifts the performance bottleneck away from throughput-oriented acceleration and toward memory bandwidth, cache efficiency, and vectorization. This makes llama.cpp a relevant proxy workload for evaluating modern CPU-based HPC systems in terms of both performance and efficiency \cite{MARQUESGARCIA2026108242, seonjin2024cpullm}.
In our evaluation, we use the benchmarking utility provided by \texttt{llama.cpp}, namely \texttt{llama-bench}. The tool enables the configuration of the LLM model, prompt and generated-token lengths, and batch size, and reports the achieved throughput in tokens per second ($tok/s$) for both the prompt-processing and text-generation phases.

The experiments are based on GPT-OSS-20B~\cite{agarwal2025gpt}, a 20-billion-parameter Mixture-of-Experts (MoE) model using FP16 weights. This model was selected to investigate CPU-based inference of an open-weight frontier model while keeping the execution time of the experiments manageable. We consider a prompt length of 2048 tokens and generate an additional 512 tokens during each inference. The batch size is varied from 1 to 512 in powers-of-two increments via the configurable \texttt{batch} and \texttt{ubatch} sizes that are kept equal to avoid pipelining and its side-effects. For each configuration, experiments are repeated five times while scaling the number of OpenMP threads from one to the maximum number of threads available on the evaluated system. The reported throughput corresponds to the arithmetic mean of the five measurements, expressed in tokens per second ($tok/s$).

%% file: results.tex
\section{Results} \label{sec:results}
In this section, we discuss the results obtained across the three evaluated platforms using the selected set of synthetic benchmarks and applications. The analysis aims to characterize the performance of the RISC-V SG2044 processor, and to assess its position relative to representative x86-64 and Arm-based competing platforms.
As a general rule, we present the results from two complementary perspectives. First, we report the raw performance or time-to-solution, depending on the application. Second, we evaluate scalability through the relative speedup, taking the single-core result as the baseline and considering the scaling up to the maximum number of available cores. The reported runs are, otherwise indicated, the best. In the totality of cases this refers to the process pinning labeled as best in Table \ref{tab:placement_policies}. 
%To assess how closely the applications follow ideal linear scaling, we also report parallel efficiency, defined as the achieved performance at each core count relative to the corresponding ideal speedup, expressed as a percentage. 
Finally, we consider the energy dimension. For benchmarks measuring wall-clock execution time on the same problem size across the evaluated architectures, the comparison is performed directly in terms of total energy consumption, expressed in kJ. This metric captures the total energy required by each platform to complete the same workload. For tests characterized by sustained performance, expressed in GFLOP/s or tok/s, including cases with different problem sizes, energy efficiency is instead measured as the ratio between the achieved performance and the energy consumed, expressed in GFLOP/J or tok/J, respectively. 

\subsection{Synthetic Benchmarks}
In this section we will present all the synthetic benchmarks results introduced above: (i) STREAM (ii) HPL (iii) HPCG and (iv) HPL-MxP.
\subsubsection{STREAM}\label{sec:res_stream}
\begin{figure*}[t]
    \centering
    % Left side: GFLOP/s Performance
    \begin{subfigure}[b]{0.48\textwidth}
        \centering
        \includegraphics[width=\linewidth]{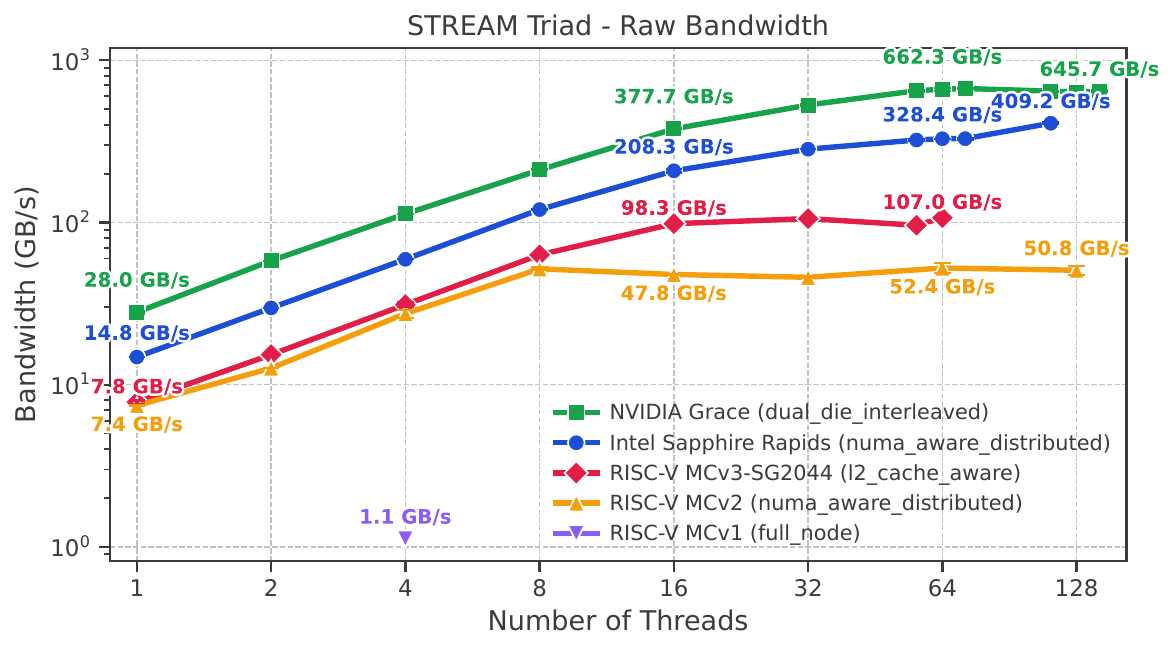}
        \caption{Sustained GB/s performance.}
        \label{fig:stream_gb_sub}
    \end{subfigure}
    \hfill % Adds flexible space between the two images
    % Right side: Relative Speedup
    \begin{subfigure}[b]{0.48\textwidth}
        \centering
        \includegraphics[width=\linewidth]{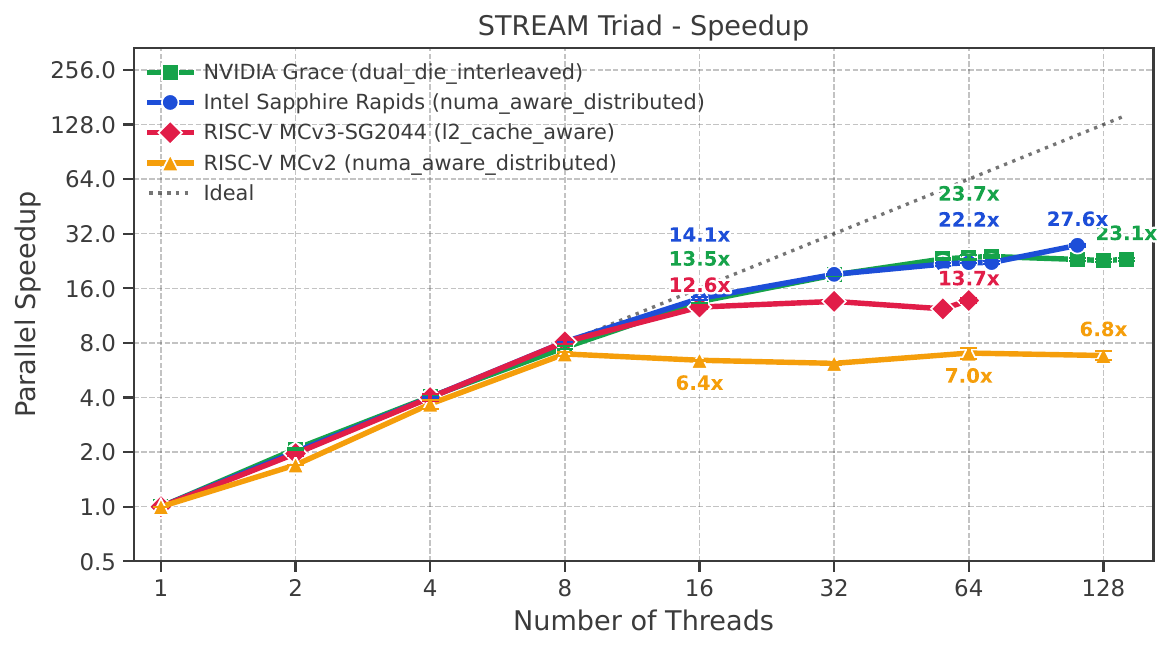}
        \caption{Relative speedup scaling.}
        \label{fig:stream_speedup_sub}
    \end{subfigure}
    
    % Main combined caption and label
    \caption{Memory bandwidth across architectures and generations of Monte Cimone. (a) shows GB/s with scaling number of OpenMP threads (here equal to the number of cores) and (b) relative speedup.}
    \label{fig:stream_combined_Scaling}
\end{figure*}

\begin{figure*}[t]
    \centering
    % Left side: GFLOP/s Performance
    \begin{subfigure}[b]{0.46\textwidth}
        \centering
        \includegraphics[width=\linewidth]{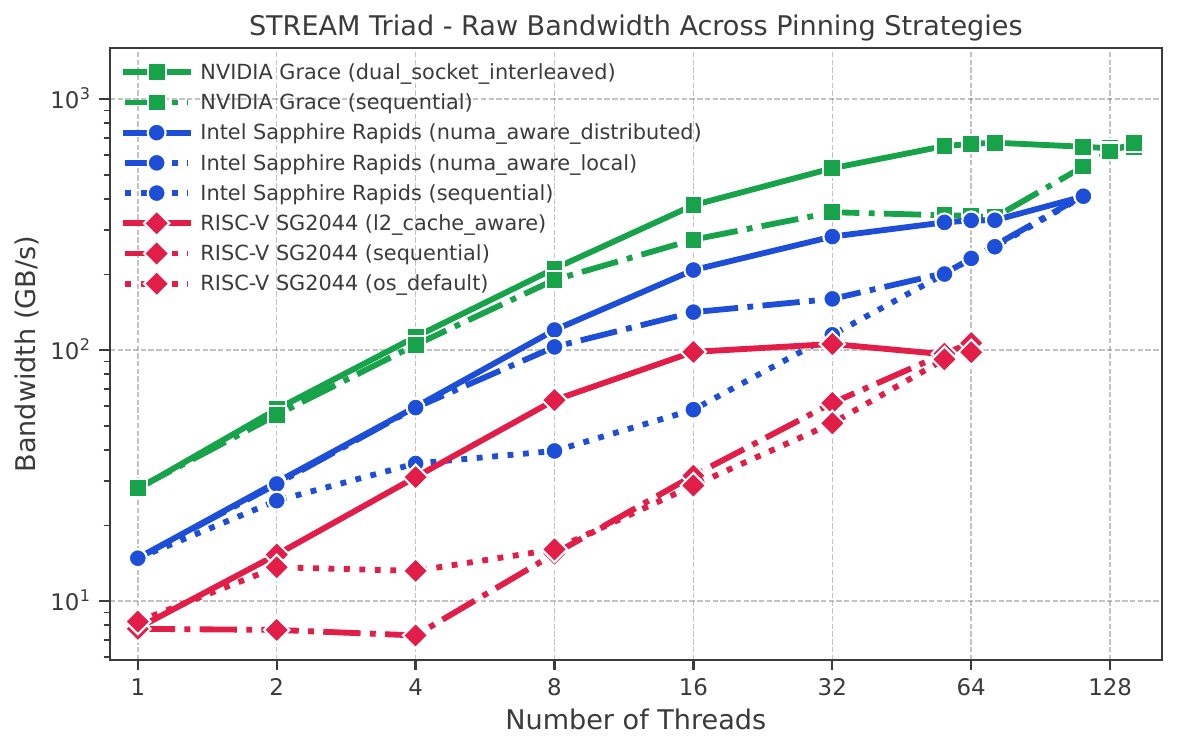}
        \caption{Sustained GB/s performance.}
        \label{fig:stream_all_pin_gb_sub}
    \end{subfigure}
    \hfill % Adds flexible space between the two images
    % Right side: Relative Speedup
    \begin{subfigure}[b]{0.46\textwidth}
        \centering
        \includegraphics[width=\linewidth]{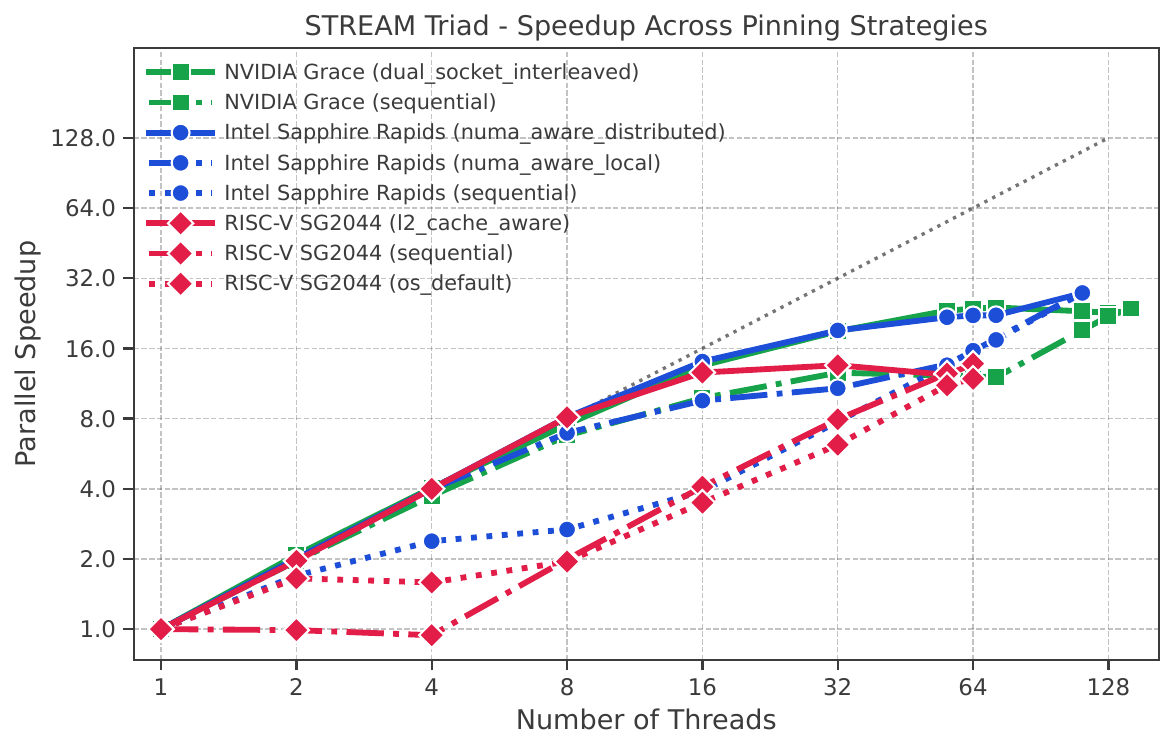}
        \caption{Relative speedup scaling.}
        \label{fig:stream_all_pin_speedup_sub}
    \end{subfigure}
    
    % Main combined caption and label
    \caption{Memory bandwidth across architectures and different pinning strategies. (a) shows GB/s with scaling number of OpenMP threads (here equal to the number of cores) and (b) relative speedup.}
    \label{fig:stream_all_pinnings}
\end{figure*}
Previous works~\cite{BrownJamieson2024_SG2042,venieri2025monte} highlighted the limitations of the memory subsystem of the previous-generation SG2042 processor. In particular, multi-core scaling rapidly reached the available memory-bandwidth ceiling: depending on the system configuration, bandwidth saturated at four cores on single-socket systems and at eight cores on dual-socket systems. This behavior is visible in Figure~\ref{fig:stream_gb_sub}, where the yellow curve represents the dual-socket SG2042 system of MCv2. Motivated by this limitation, we investigated whether the redesigned memory subsystem of the SG2044 could improve not only peak bandwidth, but also its scaling behavior.

Previous measurements on the SG2044 reported more than a twofold increase in maximum attainable memory bandwidth compared with the SG2042~\cite{Brown2025_SG2044}; however, the scaling behavior across cores was not investigated in detail. Although the SG2044 is a UMA processor, its 64 cores are organized into 16 four-core clusters, each sharing an L2 cache. This organization suggests that core placement may influence the ability to exploit the available memory bandwidth. We therefore evaluated three pinning strategies, as detailed in Section \ref{sec:pinnings}: \texttt{os\_default}, \texttt{sequential}, and \texttt{l2\_cache\_aware}. Equivalent strategies were evaluated on the Intel and NVIDIA systems by using their NUMA domains as the corresponding hardware locality units.

The results are shown in Figure~\ref{fig:stream_all_pinnings}. Distributing the workload across NUMA domains, or, for the SG2044, across core clusters, substantially improves parallel scaling compared with concentrating cores within a single locality domain. While expected on NUMA systems, the SG2044 shows that a similar effect can arise in a UMA architecture with shared-cache clusters. Its sequential configuration, shown by the red dash-dotted curve, provides little additional bandwidth once a cluster is saturated, while the unconstrained configuration performs better but does not consistently exploit the available bandwidth. The L2-cache-aware configuration achieves the best scaling by distributing active cores across the 16 clusters.
This behavior identifies the point at which the SG2044 approaches memory-bandwidth saturation. With L2-cache-aware placement, bandwidth scales nearly ideally up to 16 cores, reaching a $12.6\times$ speedup, after which gains become marginal and reach only $13.5\times$ at 64 cores (Figure~\ref{fig:stream_speedup_sub}). Compared with the SG2042, where saturation occurred at eight cores, the SG2044 shifts the onset to 16 cores while substantially increasing attainable bandwidth.

Figure~\ref{fig:stream_gb_sub} shows a comparison of the memory-bandwidth evolution across the evaluated platforms. The first generation of Monte Cimone, based on a four-core RISC-V system, achieved only 1.1~GB/s in a single node. This increased to 52.4~GB/s in MCv2, based on the SG2042, and to 107.0~GB/s in MCv3, based on the SG2044. Thus, the aggregate bandwidth increased by approximately $50\times$ between the first and second generations and by approximately $100\times$ between the first and third generations. The SG2044 therefore represents a substantial improvement in the memory performance of the RISC-V platform, although a significant gap remains with the x86-64 and Arm reference systems.

In terms of single-core bandwidth, the SG2044 reaches 7.8~GB/s, compared with 14.8~GB/s for Intel Sapphire Rapids and 28.0~GB/s for NVIDIA Grace CPU. The difference is even more pronounced when considering aggregate node bandwidth: the Intel system reaches 409.2~GB/s with 112 cores, while the NVIDIA system reaches 662.3~GB/s with 144 cores, compared with 107.0~GB/s for the 64-core SG2044. Therefore, the SG2044 remains behind the reference platforms in both per-core and aggregate bandwidth. However, absolute bandwidth alone does not capture the evolution of the memory subsystem, since the ability to scale toward the available bandwidth is equally important.

This is illustrated in Figure~\ref{fig:stream_speedup_sub}. At 16 cores, Intel Sapphire Rapids, NVIDIA Grace CPU, and SG2044 achieve speedups of $14.1\times$, $13.5\times$, and $12.6\times$, respectively, relative to single-core execution, compared with the ideal $16\times$. At their respective maximum core counts, Intel reaches $27.6\times$ at 112 cores, NVIDIA reaches $23.1\times$ at 144 cores, and SG2044 reaches $13.5\times$ at 64 cores. The comparison shows that, although the SG2044 still provides lower absolute memory bandwidth, its redesigned memory subsystem has shifted the saturation point to a significantly higher core count. Consequently, the SG2044 exhibits a scaling behavior that is much closer to that of the Intel and NVIDIA platforms than that of the previous SG2042 generation, demonstrating a substantial improvement in the ability of the RISC-V platform to exploit parallel memory bandwidth.

\subsubsection{HPL}
\begin{figure*}[t]
    \centering
    % Left side: GFLOP/s Performance
    \begin{subfigure}[b]{0.46\textwidth}
        \centering
        \includegraphics[width=\linewidth]{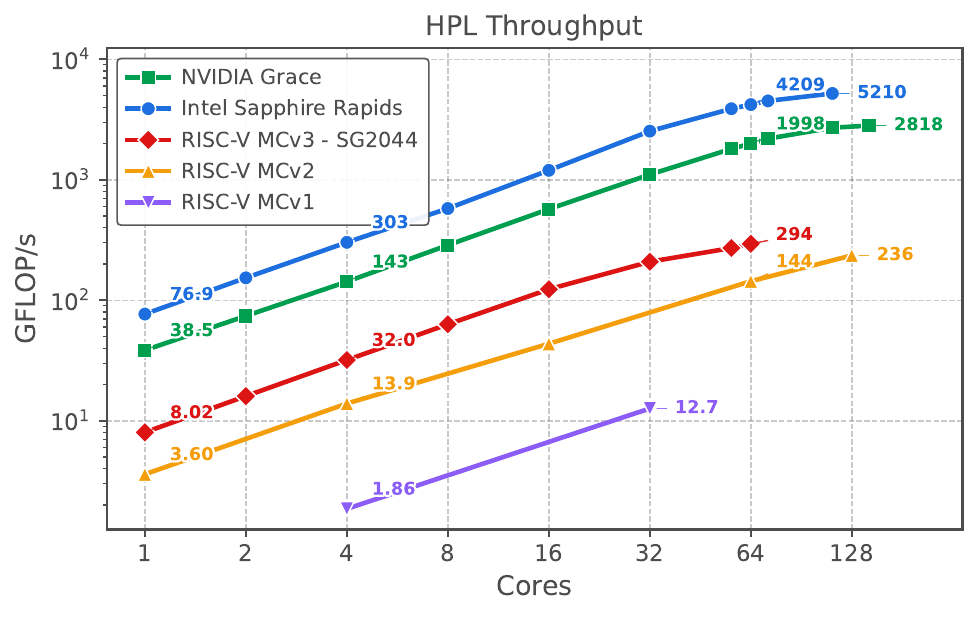}
        \caption{Sustained GFLOPS/s performance.}
        \label{fig:hpl_raw_sub}
    \end{subfigure}
    \hfill % Adds flexible space between the two images
    % Right side: Relative Speedup
    \begin{subfigure}[b]{0.46\textwidth}
        \centering
        \includegraphics[width=\linewidth]{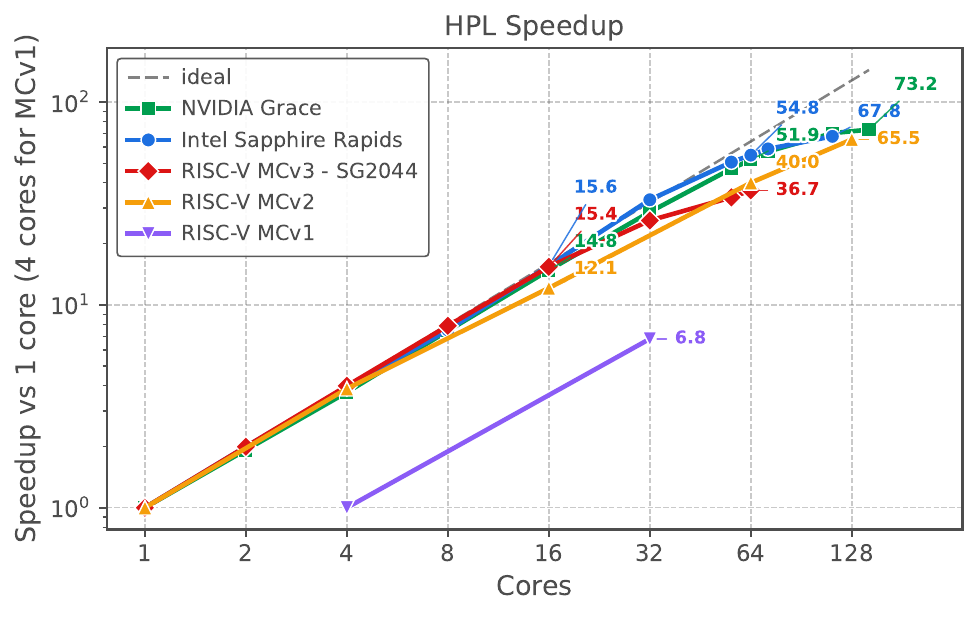}
        \caption{Relative speedup scaling.}
        \label{fig:hpl_speedup_sub}
    \end{subfigure}
    
    % Main combined caption and label
    \caption{HPL throughput in GFLOPS/s (FP64) across architectures and generations of Monte Cimone. (a) shows raw performance with scaling number of MPI processes and (b) relative speedup.}
    \label{fig:hpl_combined_scaling}
\end{figure*}
Following the evolution observed from Monte Cimone v1 to v2, the SG2044-based MCv3 further increases available per core computational power. The SG2044 reaches 8.02~GFLOP/s in single-core HPL, compared to 3.60~GFLOP/s of the SG2042 and 1.86~GFLOP/s of the processor used in MCv1. At the node level, the improvement is substantial: the SG2044 reaches 294~GFLOP/s with 64 cores, compared with 236~GFLOP/s for the SG2042-based node with 128 cores. Thus, the third-generation platform demonstrates a substantial generational improvement with higher peak HPL performance with half the number of cores of MCv2.

Figure~\ref{fig:hpl_raw_sub} places these results in the context of the complete testbed. Despite the significant improvement over previous RISC-V generations, the SG2044 remains behind the x86-64 and Arm reference platforms in absolute performance. The difference is already evident at single-core level: NVIDIA Grace and Intel Sapphire Rapids reach 38.5~GFLOP/s and 76.9~GFLOP/s, respectively, compared with 8.02~GFLOP/s for the SG2044. When normalized to frequency, the relative performance approximately follows the aggregate vector width per core: 128-bit for SG2044, four 128-bit units for Grace, and two 512-bit units for Sapphire Rapids. This indicates the importance of vector resources, although performance depends on instruction throughput, memory behavior, and frequency.

The scaling results in Figure~\ref{fig:hpl_speedup_sub} provide a different perspective. The SG2044 follows near-linear scaling up to 16 cores, after which the speedup increasingly deviates from the ideal curve. At 64 cores, the SG2044 reaches a speedup of $36\times$, compared with $40\times$ for the dual-socket SG2042. Although the HPL scaling behavior therefore remains less efficient than that of the previous-generation system at this point, the SG2044 achieves substantially higher absolute performance because of its much higher single-core capability. At 64 cores, NVIDIA Grace and Intel Sapphire Rapids reach speedups of $51.9\times$ and $54.8\times$, respectively. Their scaling also progressively departs from ideality at higher core counts, reaching $73.2\times$ at 144 cores for Grace and $67.8\times$ at 112 cores for Sapphire Rapids. The SG2042 reaches $65.5\times$ at 128 cores.
The scaling behavior highlights a trade-off between per-core compute capability and parallel efficiency. At low core counts, the wider vector resources and higher frequencies of NVIDIA and Intel yield higher single-core and aggregate HPL performance. At higher counts, scaling becomes increasingly sublinear, partly due to frequency reduction under high utilization and power constraints, particularly on Sapphire Rapids. For the SG2044, scaling degradation from approximately 16 cores is consistent with the STREAM results.

Overall, the SG2044 represents a substantial generational improvement in RISC-V compute performance. Its peak HPL performance reaches 294~GFLOP/s at 64 cores, compared with 1,998~GFLOP/s for NVIDIA Grace at the corresponding 64-core point and 4,209~GFLOP/s for Sapphire Rapids. At their respective maximum core counts, Grace reaches 2,818~GFLOP/s at 144 cores and Sapphire Rapids 5,210~GFLOP/s at 112 cores. The RISC-V platform therefore remains significantly behind the competing architectures in absolute compute throughput, but the progression from MCv1 through MCv2 to MCv3 demonstrates a clear increase in both per-core performance and single-processor computational capability.

\subsubsection{HPCG}
\begin{figure*}[t]
    \centering
    % Left side: GFLOP/s Performance
    \begin{subfigure}[b]{0.48\textwidth}
        \centering
        \includegraphics[width=\linewidth]{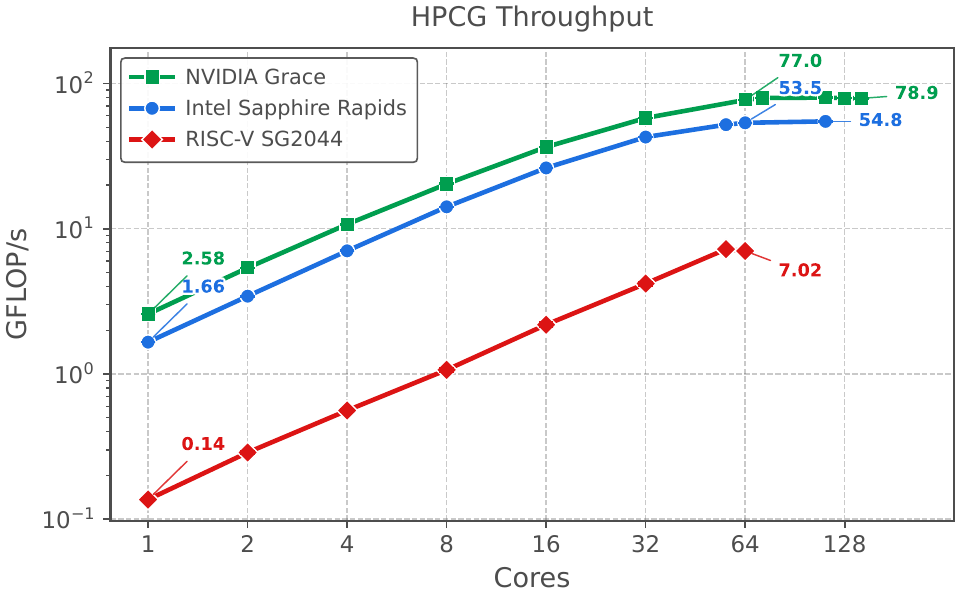}
        \caption{Strong-scaling sustained GFLOP/s performance.}
        \label{fig:hpcg_gflops_sub}
    \end{subfigure}
    \hfill % Adds flexible space between the two images
    % Right side: Relative Speedup
    \begin{subfigure}[b]{0.48\textwidth}
        \centering
        \includegraphics[width=\linewidth]{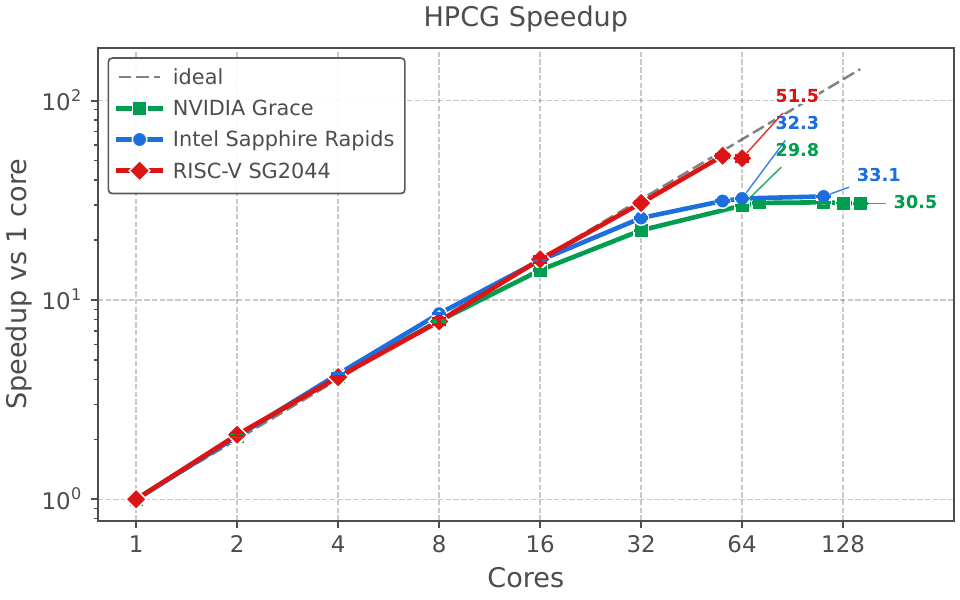}
        \caption{Relative speedup scaling.}
        \label{fig:hpcg_speedup_sub}
    \end{subfigure}
    
    % Main combined caption and label
    \caption{Strong-scaling performance results for the HPCG benchmark on the evaluated platforms as a function of the number of physical cores. (a) shows sustained performance in GFLOP/s, and (b) shows relative speedup.}
    \label{fig:hpcg_combined_scaling}
\end{figure*}
Figure~\ref{fig:hpcg_combined_scaling} shows HPCG throughput (measured in GFLOP/s) across the three evaluated architectures as a function of the core count, for a fixed 40.5~GB problem size.
At the single-core baseline, SG2044 reaches 0.137~GFLOP/s, making it $18.9\times$ slower than Grace at 2.58~GFLOP/s, and $12.1\times$ slower than Sapphire Rapids at 1.66~GFLOP/s.
Grace is the first to deviate from ideal scaling, delivering 36.50~GFLOP/s ($14.1\times$) at 16 cores, while Sapphire Rapids is close to ideal at 26.19~GFLOP/s ($15.8\times$ speedup), and SG2044 tracks the ideal scaling line at 2.18~GFLOP/s ($16.0\times$ speedup).
By 32 cores both Arm and x86 platforms experience a noticeable decline in speedup scaling:
Grace reaches 57.79~GFLOP/s ($22.4\times$ speedup) and Sapphire Rapids reaches 42.72~GFLOP/s ($25.8\times$, 80.6\% efficiency).
In contrast, SG2044 maintains robust scaling with 4.19~GFLOP/s ($30.7\times$ speedup).
At 64 cores, while the absolute performance hierarchy remains unchanged (76.96~GFLOP/s on Grace, 53.54~GFLOP/s on Sapphire Rapids, and 7.02~GFLOP/s on SG2044), the scaling behaviour completely reverses.
Grace and Sapphire Rapids retain only a fraction of their ideal scaling factor, achieving $29.8\times$ and $32.3\times$ speedup, respectively.
Conversely, SG2044 sustains a $51.5\times$ speedup, after peaking at 7.22~GFLOP/s ($52.9\times$, 94.5\% efficiency) on 56 cores.
The performance gap therefore narrows from $18.9\times$ and $12.1\times$ at one core to $11.0\times$ and $7.6\times$ at 64 cores.
At the largest core count of each machine the two server platforms are saturated.
Grace plateaus from 72 cores onward (79.34~GFLOP/s), peaking at 79.87~GFLOP/s on 112 cores and settling at 78.87~GFLOP/s on all 144 cores ($30.5\times$, 21.2\% efficiency).
Sapphire Rapids reaches 54.79~GFLOP/s on 112 cores ($33.1\times$ speedup), only 2.3\% above its own 64-core result.
SG2044 ends at its full 64 cores with 7.02~GFLOP/s ($51.4\times$ speedup), which is $11.2\times$ below Grace at 144 cores and $7.8\times$ below Sapphire Rapids at 112 cores in absolute terms.

\subsubsection{HPL-MxP}
% Grafici saphhre pronti
% Grafici neoverse pronti
% Grafici sg2044  pronti
% Grafici power pronti
\begin{figure*}[t]
    \centering
    \begin{subfigure}[b]{0.46\textwidth}
        \centering
        \includegraphics[width=\linewidth]{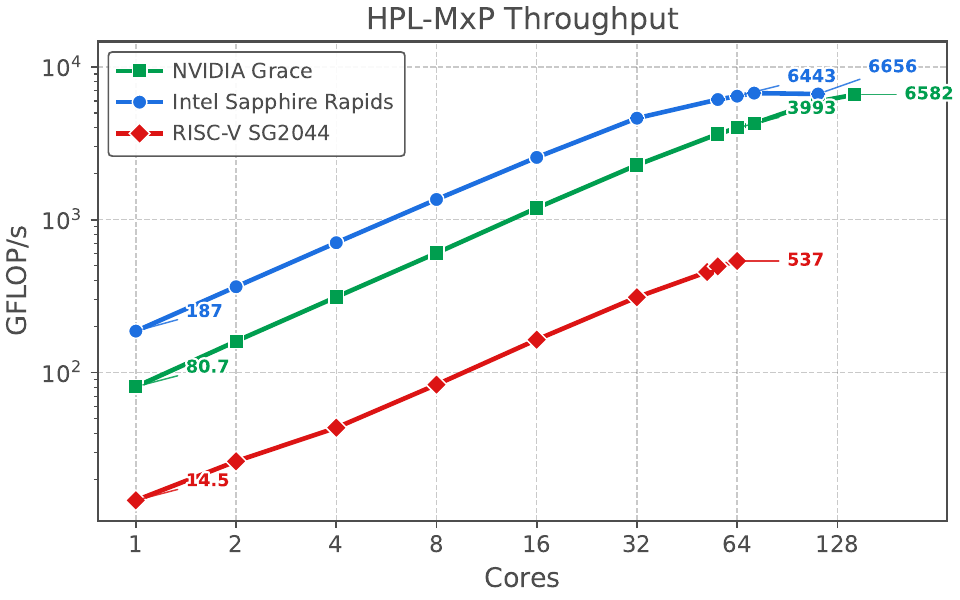}
        \caption{Sustained GFLOPS/s performance.}
        \label{fig:hpl_mxp_glops}
    \end{subfigure}
    \hfill
    \begin{subfigure}[b]{0.46\textwidth}
        \centering
        \includegraphics[width=\linewidth]{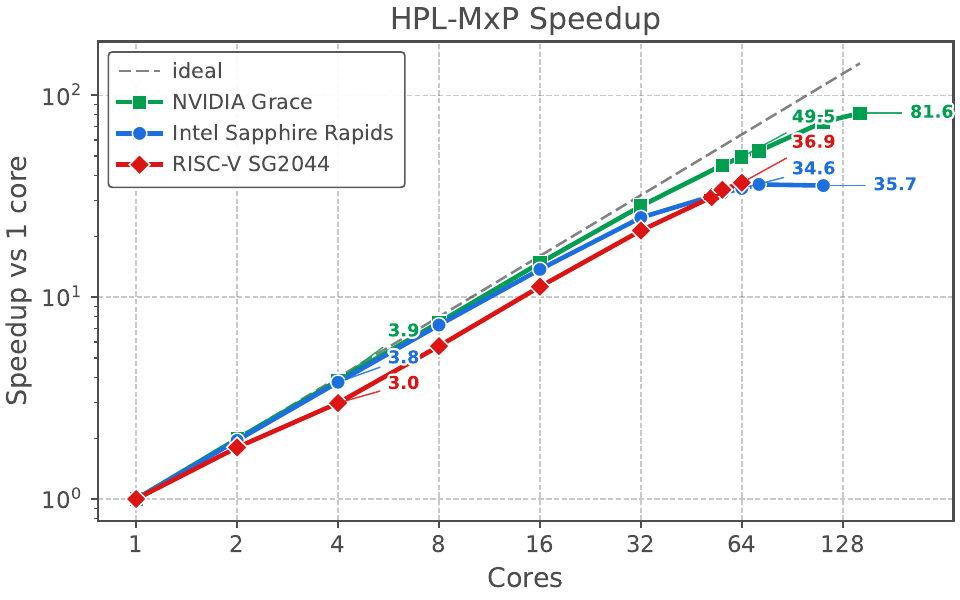}
        \caption{Relative speedup scaling.}
        \label{fig:hpl_mxp_scaling}
    \end{subfigure}

    \caption{HPL MxP throughput in GFLOPS/s (mixed precision) across different architectures. (a) shows raw performance (GFLOP/s) on x-axes over scaling number cores (y-axes) and (b) relative speedup on x-axes over scaling number cores (y-axes).}
    \label{fig:hpl_mxp_combined_scaling}
\end{figure*}

% From Figure \ref{fig:hpl_mxp_combined_scaling} we can see the results of HPL mixed precision throughput across the three evaluated architectures as a function of the core count, for a fixed memory utilization (>70\%).
% At the single-core baseline, SG2044 delivers $14,46$ GFLOP/s, corresponding to a $12,90$ performance deficit relative to Sapphire Rapids, which reaches $186,54$ GFLOP/s, and a $5,58$ deficit compared with Grace, at $80,7$ GFLOP/s. 

% % Done
% SG2044 moves slightly away at the core count to $4$ from the ideal speedup as depicted in \ref{fig:hpl_mxp_scaling}. Sapphire Rapids speedup increase up to core count $56$, before 
% departing, reaching $6124.67$ GFLOP/s ($32,83$×), whereas SG2044 continues to follow near-ideal scaling, reaching at 64 core $536,54$ GFLOP/s ($36,897$ × speedup). Grace remains close to the ideal trend until $112$ cores reaching $5953,04$ GFLOP/s ($73,79 $× speedup).  At the maximum core count available on each system, the two server platforms reach saturation. Grace levels off from  cores onward, attaining a peak of  GFLOP/s before settling at  GFLOP/s across all  cores (×, \% efficiency). Sapphire Rapids reaches  GFLOP/s at  cores (× speedup), just \% above its -core result. SG2044 reaches its maximum configuration of  cores with  GFLOP/s (× speedup), remaining × and × behind Grace and Sapphire Rapids, respectively, in absolute performance. Overall best speedup is reached by Grace platform followed by RISC-V that overtook Sapphire Rapids at $64$ cores.
From Figure \ref{fig:hpl_mxp_combined_scaling} we can see the results of HPL mixed precision throughput (\ref{fig:hpl_mxp_glops}) and relative speedup (\ref{fig:hpl_mxp_scaling}) across the three evaluated architectures as a function of the core count, for a fixed memory utilization (>70\%).
At the single-core baseline, SG2044 delivers $14.54$ GFLOP/s, corresponding to a $12.83\times$ performance deficit relative to Sapphire Rapids, which reaches $186.54$ GFLOP/s, and a $5.55\times$ deficit compared with Grace, at $80.68$ GFLOP/s. 
SG2044 moves slightly away at the core count to $4$ from the ideal speedup as depicted in \ref{fig:hpl_mxp_scaling} and continues to follow near-ideal scaling, reaching $536.55$ GFLOP/s ($36.90\times$ speedup) at $64$ cores. Sapphire Rapids speedup increases up to core count $56$ before departing, reaching $6124.67$ GFLOP/s ($32.83\times$). Grace remains close to the ideal trend until $112$ cores, reaching $5953.04$ GFLOP/s ($73.79\times$ speedup). At the maximum core count available on each system, the two server platforms reach saturation. Sapphire Rapids levels off from $72$ cores onward, attaining a peak of $6747.62$ GFLOP/s before settling at $6656.47$ GFLOP/s across all $112$ cores ($35.69\times$). Grace reaches $6582.40$ GFLOP/s at $144$ cores ($81.59\times$ speedup), just $10.57\%$ above its $112$-core result. SG2044 reaches its maximum configuration of $64$ cores with $536.55$ GFLOP/s ($36.90\times$ speedup), remaining $12.27\times$ and $12.41\times$ behind Grace and Sapphire Rapids, respectively, in absolute performance. Overall best speedup is reached by the Grace platform, followed by RISC-V, which overtook Sapphire Rapids at $64$ cores.
%TODO: Confronto con HPL
\subsection{Applications}
In this section we will present all the application results introduced above: (i) iPIC3D, (ii) PLUTO, (iii) OpenFOAM and (iv) llama-bench. 
\subsubsection{iPIC3D}

%\vspace{1.5em}
\begin{figure*}[t]
    \centering
    % Requires \usepackage{subcaption} in your preamble
    
    \begin{subfigure}[b]{0.48\textwidth}
        \centering
        \includegraphics[width=\linewidth]{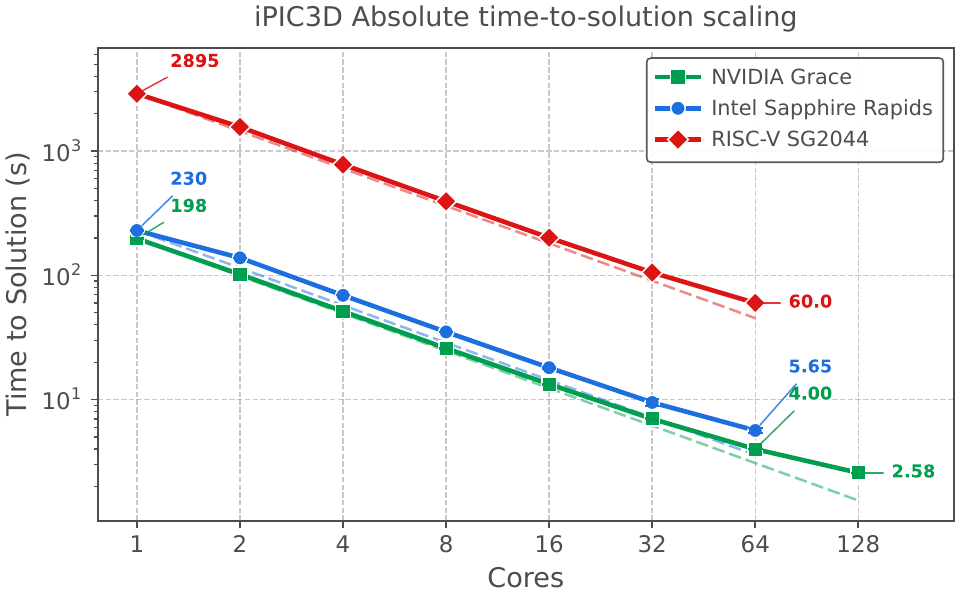}
        \caption{Absolute time-to-solution scaling.}
        \label{fig:ipic3d_time_to_solution}
    \end{subfigure}
    \hfill
    \begin{subfigure}[b]{0.48\textwidth}
        \centering
        \includegraphics[width=\linewidth]{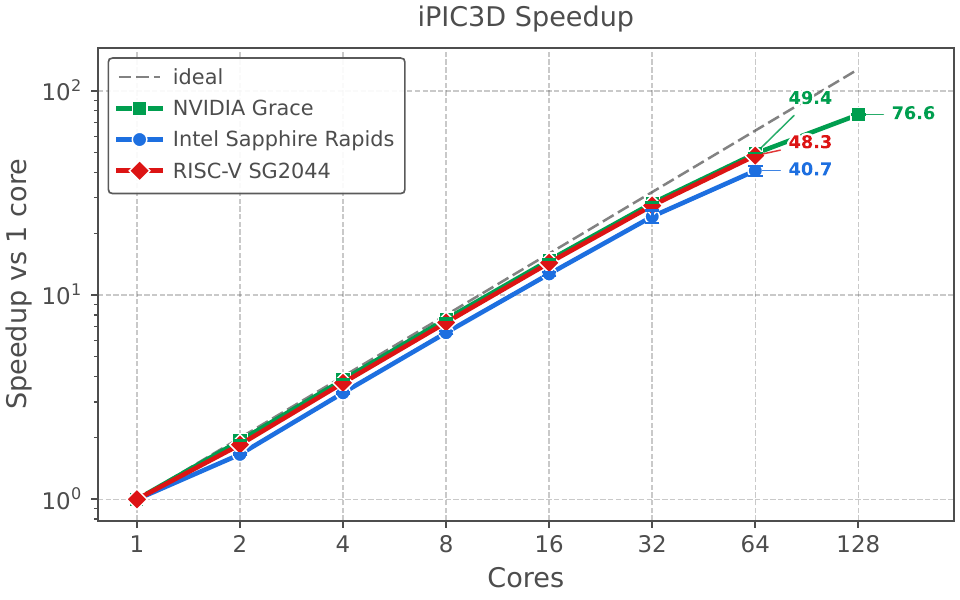}
        \caption{Relative speedup scaling.}
        \label{fig:ipic3d_speedup_scaling}
    \end{subfigure}
    \caption{Strong-scaling results for iPIC3D on the evaluated platforms as a function of the number of physical cores: (a) absolute time-to-solution, (b) relative speedup.}
    \label{fig:ipic3d_scaling_comparison}
\end{figure*} 
Figure~\ref{fig:ipic3d_time_to_solution} shows iPIC3D's time-to-solution as a function of active core count on the three architectures.
At the single-core baseline, SG2044 requires 2{,}895.2~s to complete the simulation, rendering it $14.7\times$ slower than Grace (197.6~s) and $12.6\times$ slower than Sapphire Rapids (229.6~s).
As execution scales to 8 cores, Grace and SG2044 demonstrate near-ideal scaling, achieving speedups of $7.6\times$ and $7.3\times$ respectively, while Sapphire Rapids begins to trail with a $6.5\times$ speedup.
This scaling hierarchy persists through 16 cores, where Grace reaches a $14.8\times$ speedup, SG2044 reaches $14.4\times$, and Sapphire Rapids yields $12.7\times$, and diverges further at 32 cores, yielding speedups of $28.2\times$, $27.5\times$, and $24.2\times$, respectively.
At 64 cores, which represents the maximum core count measured on both SG2044 and Sapphire Rapids, the absolute time-to-solution drops to 3.998~s on Grace, 5.647~s on Sapphire Rapids, and 59.967~s on SG2044.
At this stage, SG2044 achieves a $48.3\times$ speedup over its single-core baseline, closely tracking the $49.4\times$ speedup observed on Grace, whereas Sapphire Rapids lags significantly with a $40.7\times$ speedup.
Despite this competitive relative scaling, SG2044 remains $15.0\times$ slower than Grace and $10.6\times$ slower than Sapphire Rapids in absolute terms.
% Notably, unlike the behavior observed in HPCG and PLUTO, SG2044's robust parallel scaling does not translate into a narrower absolute performance gap against Grace at higher core counts: the gap stays essentially constant, from $14.7\times$ at one core to $15.0\times$ at 64 cores, whereas against Sapphire Rapids it narrows from $12.6\times$ to $10.6\times$.
% The memory-bound, particle-in-cell kernels of iPIC3D heavily favor the superior per-core memory bandwidth of Grace throughout the entire scaling sweep.
Ultimately, Grace alone is able to effectively scale further to 128 cores, achieving a final time-to-solution of 2.578~s (a $76.6\times$ overall speedup).

\subsubsection{PLUTO}

\begin{figure*}[t]
    \centering
    % Requires \usepackage{subcaption} in your preamble
    
    \begin{subfigure}[b]{0.48\textwidth}
        \centering
        \includegraphics[width=\linewidth]{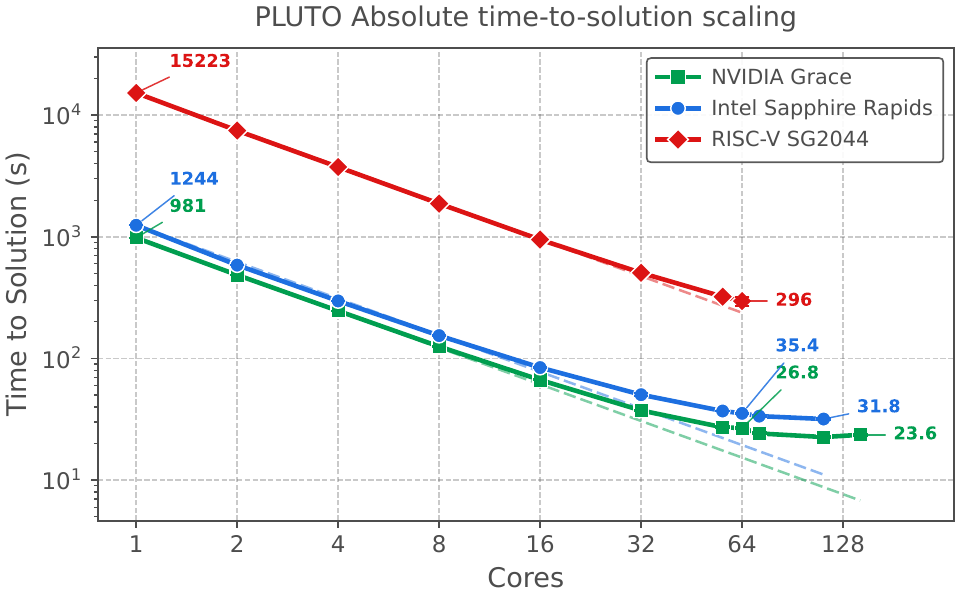}
        \caption{Absolute time-to-solution scaling.}
        \label{fig:pluto_time_to_solution}
    \end{subfigure}
    \hfill
    \begin{subfigure}[b]{0.48\textwidth}
        \centering
        \includegraphics[width=\linewidth]{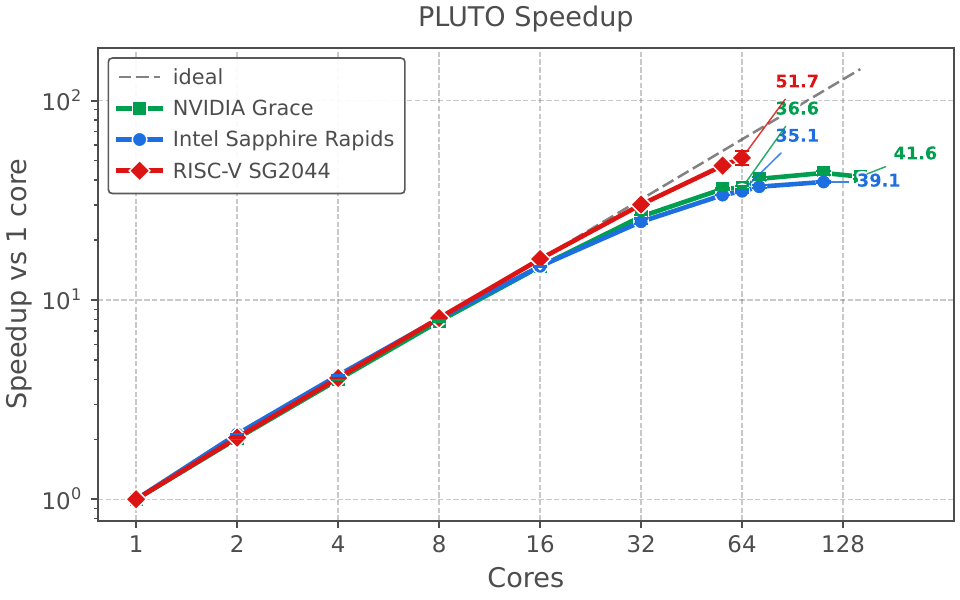}
        \caption{Relative speedup scaling.}
        \label{fig:pluto_speedup_scaling}
    \end{subfigure}
    
    \caption{Strong-scaling results for PLUTO on the evaluated platforms as a function of the number of physical cores: (a) absolute time-to-solution, (b) relative speedup.}
    
    \label{fig:pluto_scaling_comparison}
\end{figure*}

Figure~\ref{fig:pluto_time_to_solution} shows PLUTO's time-to-solution as a function of core count on the three architectures.
On a single core SG2044 takes 15{,}223~s, $15.5\times$ slower than Grace (980.8~s) and $12.2\times$ slower than Sapphire Rapids (1{,}244~s).
Up to 16 cores all three scale close to ideal, with Grace and Sapphire Rapids the first to fall behind, both at roughly 92\% efficiency (92.0\% and 92.3\%, respectively), while SG2044 is still at 100.3\%.
By 32 cores the gap widens: SG2044 still scales well at $30.1\times$ (94.1\% efficiency), while Grace reaches $26.2\times$ (82.0\%) and Sapphire Rapids only $24.7\times$ (77.1\%).
At 64 cores, SG2044's full node, the absolute ranking is unchanged (26.8~s on Grace, 35.4~s on Sapphire Rapids, 296~s on SG2044), but efficiency is reversed: SG2044 retains 80.3\% ($51.4\times$), against 57.2\% ($36.6\times$) on Grace and just 54.9\% ($35.1\times$) on Sapphire Rapids.
The absolute gap therefore narrows from $15.5\times$ and $12.2\times$ at one core to $11.1\times$ and $8.4\times$ at 64 cores.
At their largest core counts, both x86 and Arm platforms are well past this point: Grace peaks at 112 cores (22.6~s, $43.4\times$, 38.7\% efficiency) and settles slightly at 144 cores (23.6~s, $41.6\times$, 28.9\%), while Sapphire Rapids reaches 31.8~s at 112 cores ($39.1\times$, 34.9\%).
SG2044 ends at its full 64 cores with 296~s, $9.3\times$ slower than Sapphire Rapids at 112 cores and $12.6\times$ slower than Grace at 144 cores ($13.1\times$ against Grace's 112-core peak) in absolute terms, but at more than twice their parallel efficiency.

\subsubsection{OpenFOAM}
\begin{figure*}[t]
    \centering
    % Requires \usepackage{subcaption} in your preamble
    
    \begin{subfigure}[b]{0.48\textwidth}
        \centering
        \includegraphics[width=\linewidth]{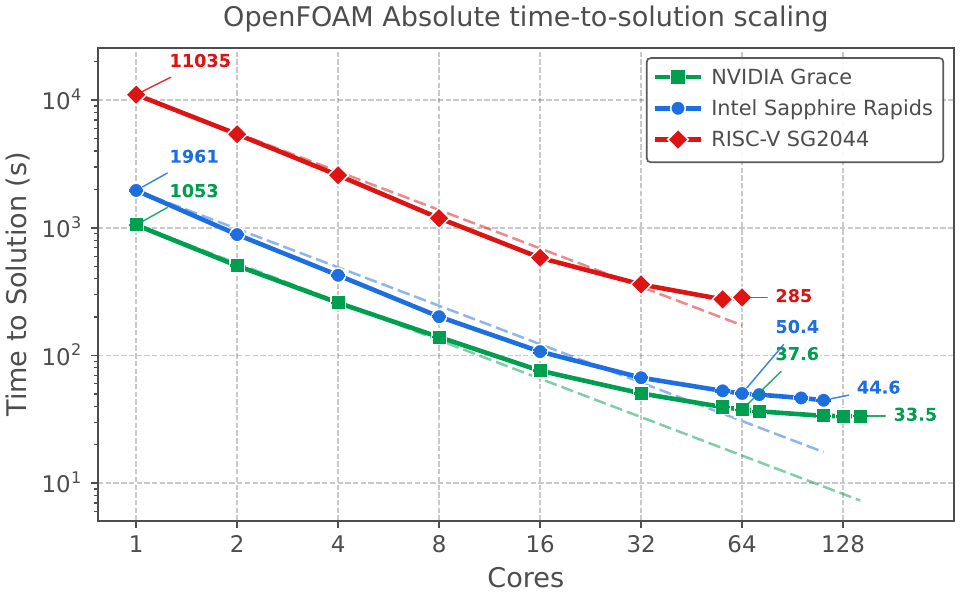}
        \caption{Absolute time-to-solution scaling.}
        \label{fig:openfoam_time_to_solution}
    \end{subfigure}
    \hfill
    \begin{subfigure}[b]{0.48\textwidth}
        \centering
        \includegraphics[width=\linewidth]{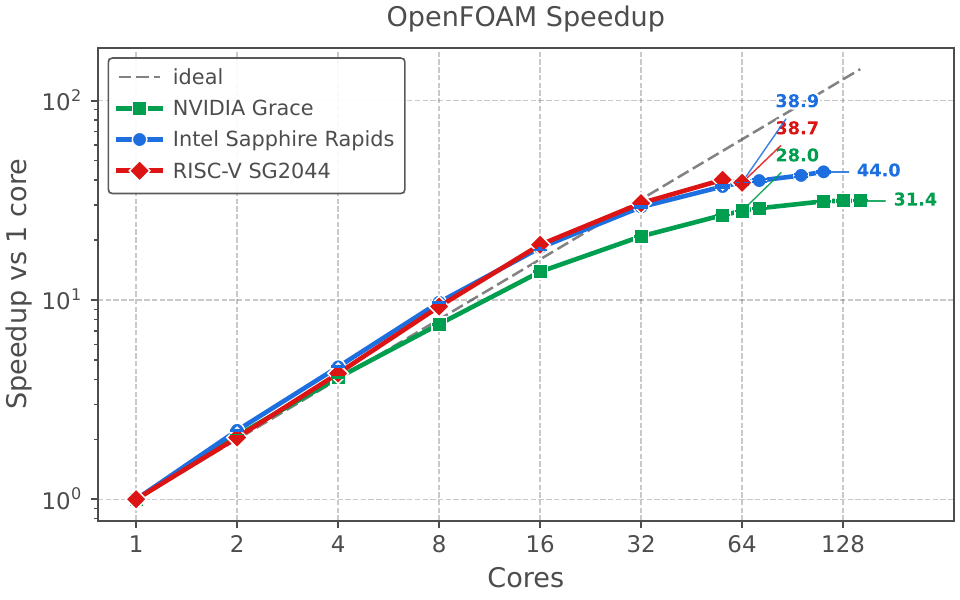}
        \caption{Relative speedup scaling.}
        \label{fig:openfoam_speedup_scaling}
    \end{subfigure}
    \hfill

    \caption{Strong-scaling results for OpenFOAM on the evaluated platforms as a function of the number of physical cores: (a) absolute time-to-solution, (b) relative speedup.}

    \label{fig:openfoam_scaling_comparison}
\end{figure*}
Figure~\ref{fig:openfoam_time_to_solution} shows OpenFOAM strong scaling on the 6M-cell Windsor body mesh.
On a single core SG2044 takes 11{,}034.8~s. It is $10.5\times$ slower than Grace, at 1{,}053.2~s, and $5.63\times$ slower than Sapphire Rapids, at 1{,}961.5~s.
Up to 8 cores Grace scales close to linearly, at $7.56\times$ speedup, while Sapphire Rapids and SG2044 scale superlinearly, reaching $9.74\times$ and $9.27\times$ respectively (121.8\% and 115.9\% efficiency). We attribute this superlinear behaviour to a cache effect: as the core count grows, the per-core subdomain shrinks and increasingly fits into the cache hierarchy, raising the effective per-core bandwidth above the single-core baseline.
From 16 cores Grace falls below the ideal line (75.97~s and 86.6\% efficiency at 16 cores, 50.47~s and 65.2\% at 32), while Sapphire Rapids and SG2044 stay close to or above it (107.23~s and 114.3\% for Sapphire Rapids, 582.46~s and 118.4\% for SG2044 at 16 cores; 66.96~s and 91.5\%, 360.23~s and 95.7\% at 32).
At 64 cores the ranking in absolute performance does not change: 37.64~s on Grace, 50.38~s on Sapphire Rapids, and 284.84~s on SG2044.
In parallel efficiency, however, Grace ranks last.
Grace retains only $27.98\times$ of the ideal factor of 64, that is 43.7\% efficiency; Sapphire Rapids and SG2044 are essentially tied, at $38.93\times$ (60.8\%) and $38.74\times$ (60.5\%) respectively.
SG2044's fastest run is in fact at 56 cores (274.95~s, $40.13\times$), after which occupying the full node yields a slight slowdown.
The gap to Grace therefore narrows from $10.5\times$ at one core to $7.57\times$ at 64 cores, while the gap to Sapphire Rapids stays essentially unchanged, from $5.63\times$ to $5.65\times$.
At the largest core count of each machine the x86 and Arm platforms are saturated.
Grace flattens beyond 128 cores, ending at 33.51~s on all 144 cores ($31.43\times$, 21.8\% efficiency).
Sapphire Rapids reaches 44.61~s on 112 cores ($43.97\times$, 39.3\% efficiency), 11.5\% below its own 64-core result.
SG2044, whose full node is the 64-core result above, is thus $8.50\times$ slower than Grace at 144 cores and $6.39\times$ slower than Sapphire Rapids at 112 cores in absolute terms, but with $2.77\times$ and $1.54\times$ their parallel efficiency, respectively.

\subsubsection{llama.cpp--llama-bench}

\begin{figure*}[t]
    \centering
    % Left side: GFLOP/s Performance
    \begin{subfigure}[b]{0.48\textwidth}
        \centering
        \includegraphics[width=\linewidth]{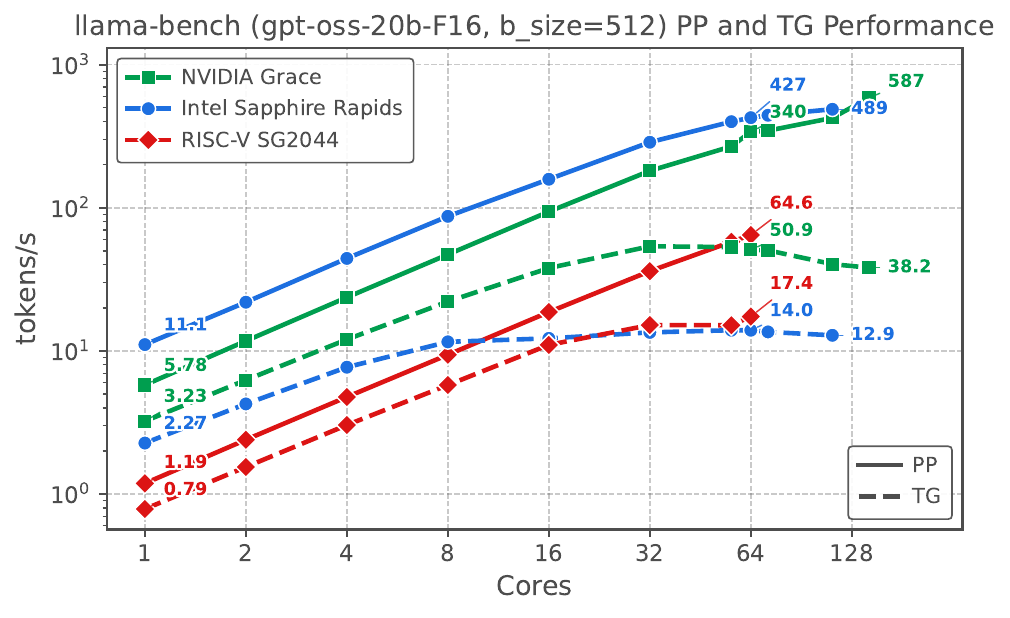}
        \caption{Prefill (PP) and decode (TG) sustained tok/s scaling.}
        \label{fig:llama_tok_sub}
    \end{subfigure}
    \hfill % Adds flexible space between the two images
    % Right side: Relative Speedup
    \begin{subfigure}[b]{0.47\textwidth}
        \centering
        \includegraphics[width=\linewidth]{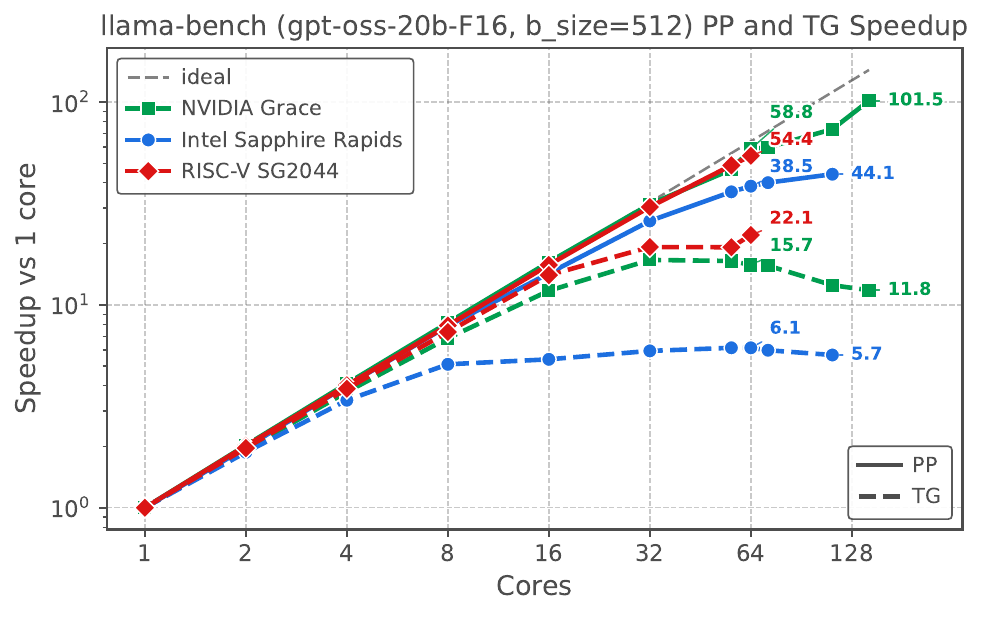}
        \caption{Prefill (PP) and decode (TG) relative speedup scaling.}
        \label{fig:llama_speedup_sub}
    \end{subfigure}

    % Main combined caption and label
    \caption{llama-bench scaling results obtained with gpt-oss-20b-F16 model and batch size equal to 512}
    \label{fig:llama_combined_scaling}
\end{figure*}

Finally, we report the LLM inference results. The evaluated configurations considered both different OpenMP thread-pinning strategies and batch sizes; here, we report the configurations that generally provided the best performance and scaling for each platform. Figure~\ref{fig:llama_combined_scaling} reports the results obtained with NUMA- and L2-cache-aware thread pinning and a batch size of 512. In general, the difference in computational capability between the SG2044 and the two comparison platforms is more pronounced during Prompt Processing (PP), whereas the performance gap is reduced during Text Generation (TG) and, in some cases, the SG2044 surpasses the Intel platform. At one core, as shown in Figure~\ref{fig:llama_tok_sub}, the SG2044 achieves 0.79~tok/s in TG and 1.19~tok/s in PP, compared with 2.27~tok/s and 11.10~tok/s for Intel Sapphire Rapids, and 3.23~tok/s and 5.78~tok/s for NVIDIA Grace, respectively. The substantially higher vector-computing capability of Sapphire Rapids translates into the highest PP throughput, whereas Grace, with higher memory bandwidth, achieves the highest TG throughput. This difference is consistent with the distinct characteristics of the two inference phases: PP is primarily compute-bound, while TG is more constrained by memory bandwidth.

For PP, the SG2044 scales close to ideally up to 64 cores, reaching 64.6~tok/s and a $54.4\times$ speedup over its single-core performance. NVIDIA Grace achieves both higher throughput, at 340~tok/s, and slightly higher relative scaling, with a $58.8\times$ speedup. Intel Sapphire Rapids reaches the highest absolute throughput at 64 cores, with 427~tok/s, but exhibits lower relative scaling, reaching a $38.5\times$ speedup. Beyond 64 cores, the difference in scaling efficiency becomes more pronounced: Sapphire Rapids reaches 489~tok/s at 112 cores, corresponding to a $44.1\times$ speedup over one core, while NVIDIA Grace continues to scale more effectively, reaching 587~tok/s at 144 cores and a $101.5\times$ speedup.

Considering TG, all platforms exhibit a stronger degradation of per-core performance as the core count increases. NVIDIA Grace achieves the highest absolute throughput, whereas the SG2044 provides the highest relative scaling among the evaluated platforms. At 64 cores, the SG2044 reaches 17.4~tok/s, corresponding to a $22.1\times$ speedup over one core. NVIDIA Grace reaches 50.9~tok/s, with a $15.7\times$ speedup, while Intel Sapphire Rapids reaches 14.0~tok/s and a $6.1\times$ speedup. These results illustrate the different scaling characteristics of the two inference phases. In the compute-bound PP phase, the SG2044 has substantially lower single-core performance than the two comparison platforms, reflecting its lower computational throughput, but maintains good scaling with increasing core count. In the more memory-bound TG phase, the SG2044 exhibits a more favorable balance between per-core compute capability and available memory bandwidth, resulting in higher scaling efficiency and, at 64 cores, higher absolute throughput than the Intel Sapphire Rapids system.

\subsection{Energy efficiency}
\begin{figure*}[t]
    \centering
    \setlength{\tabcolsep}{2pt}
    \captionsetup[subfigure]{font=footnotesize}
    \begin{tabular}{cc}
        \begin{subfigure}[t]{0.47\textwidth}
            \centering
            \includegraphics[width=\linewidth]{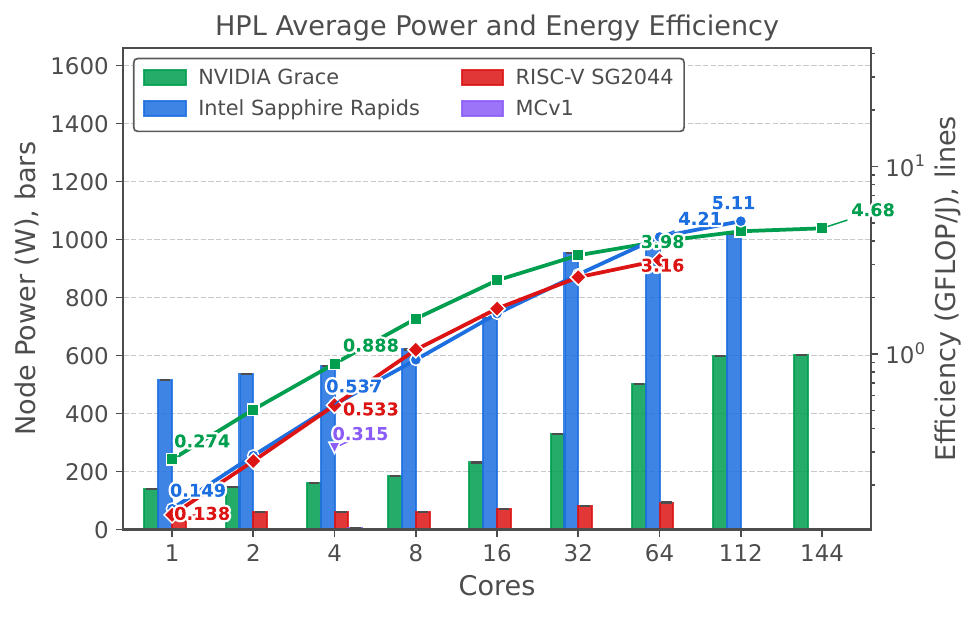}
            \caption{HPL}
            \label{fig:hpl_power_efficiency}
        \end{subfigure}
        &
        \begin{subfigure}[t]{0.47\textwidth}
            \centering
            \includegraphics[width=\linewidth]{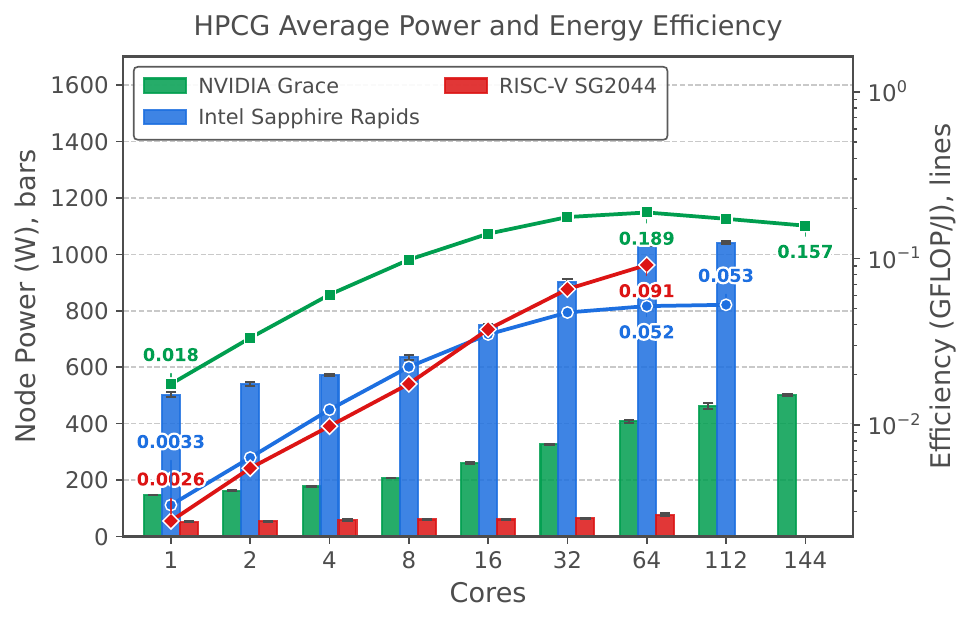}
            \caption{HPCG}
            \label{fig:hpcg_power_efficiency}
        \end{subfigure}
        \\[-0.1em]
        \begin{subfigure}[t]{0.47\textwidth}
            \centering
            \includegraphics[width=\linewidth]{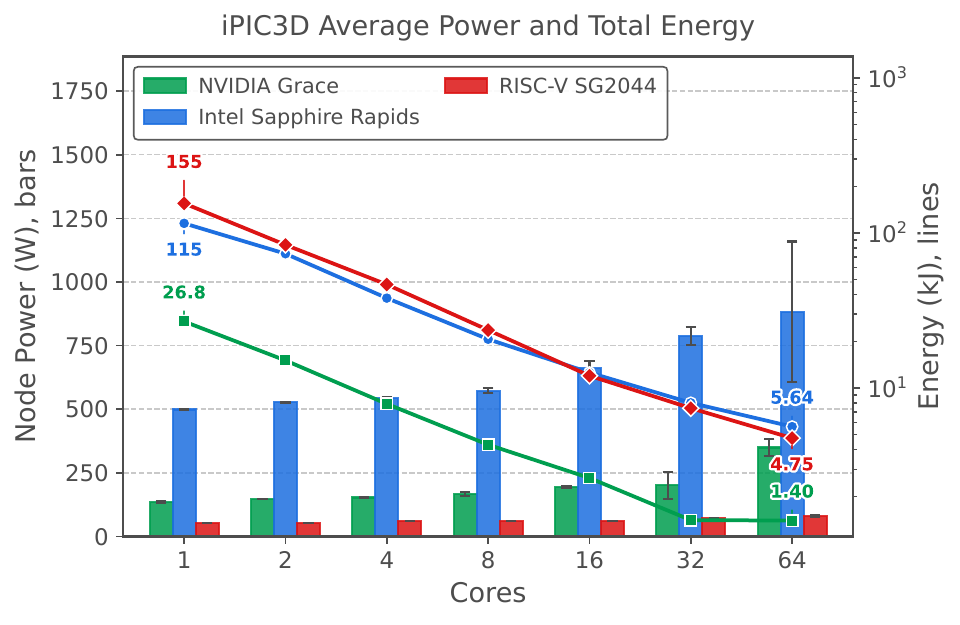}
            \caption{iPIC3D}
            \label{fig:ipic3d_power_energy}
        \end{subfigure}
        &
        \begin{subfigure}[t]{0.47\textwidth}
            \centering
            \includegraphics[width=\linewidth]{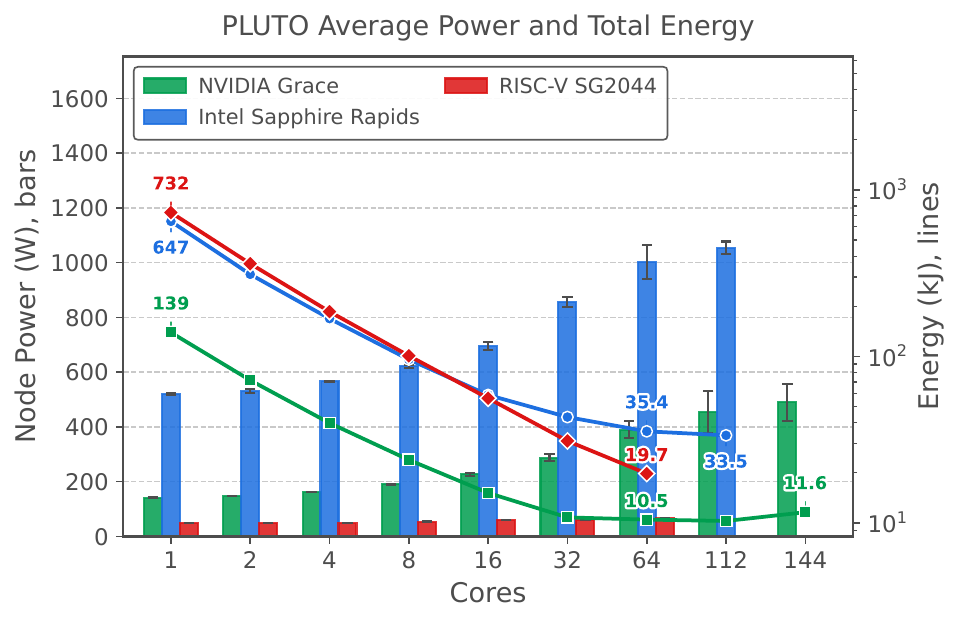}
            \caption{PLUTO}
            \label{fig:pluto_power_energy}
        \end{subfigure}
        \\[-0.1em]
        \begin{subfigure}[t]{0.47\textwidth}
            \centering
            \includegraphics[width=\linewidth]{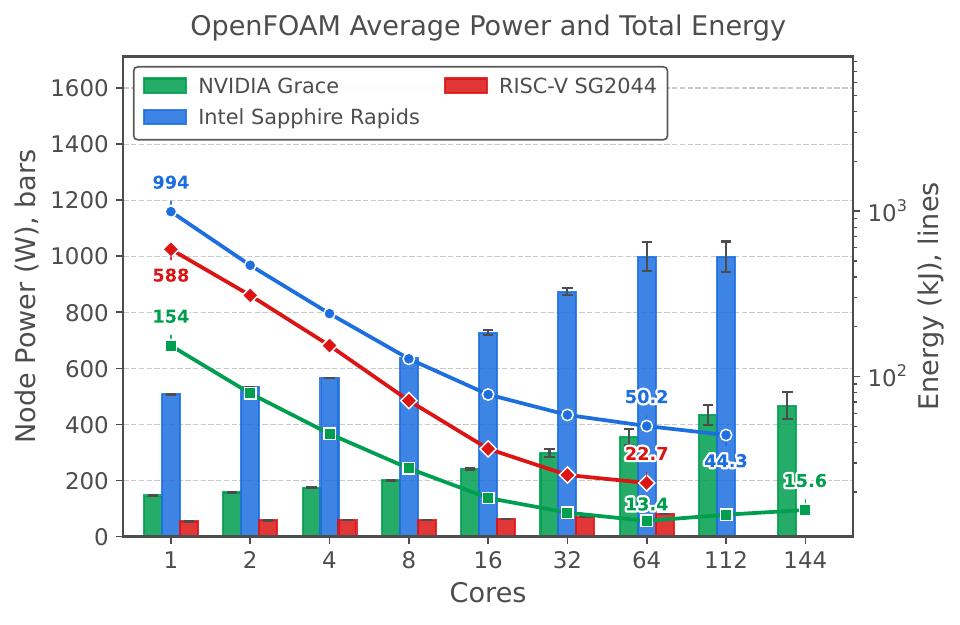}
            \caption{OpenFOAM}
            \label{fig:openfoam_power_energy}
        \end{subfigure}
        &
        \begin{subfigure}[t]{0.47\textwidth}
            \centering
            \includegraphics[width=\linewidth]{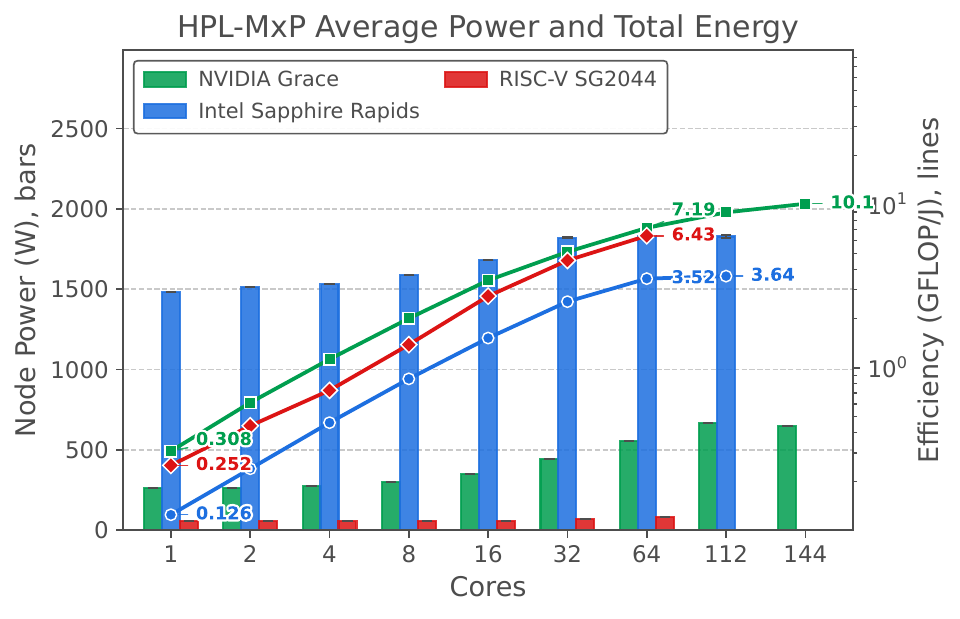}
            \caption{HPL MxP}
            \label{fig:hpl_mxp_power_efficiency}
        \end{subfigure}
        \\[-0.1em]
        \begin{subfigure}[t]{0.47\textwidth}
            \centering
            \includegraphics[width=\linewidth]{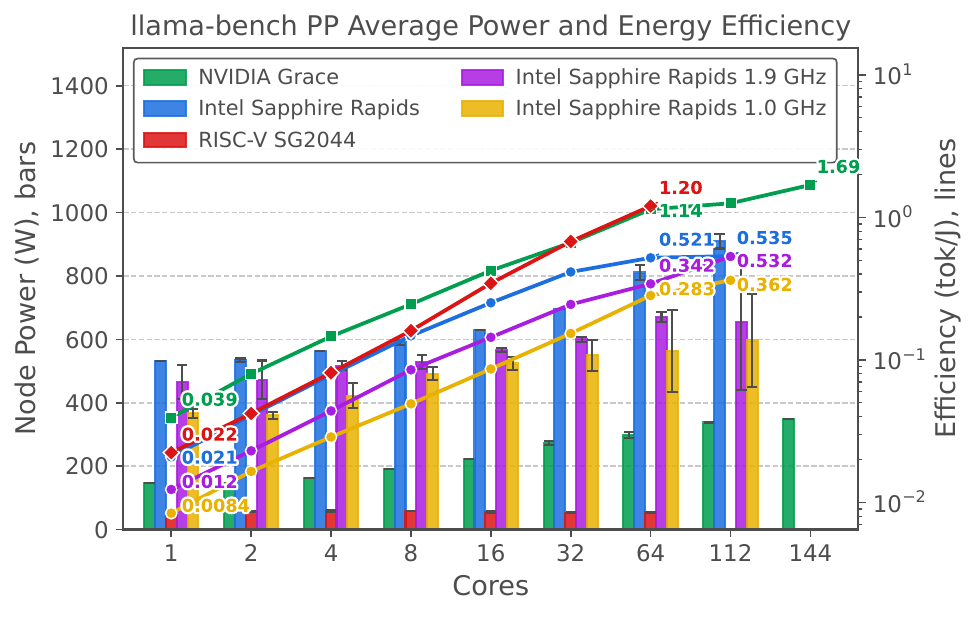}
            \caption{llama-bench Prompt-Processing}
            \label{fig:llama_power_efficiency_pp}
        \end{subfigure}
        &
        \begin{subfigure}[t]{0.47\textwidth}
            \centering
            \includegraphics[width=\linewidth]{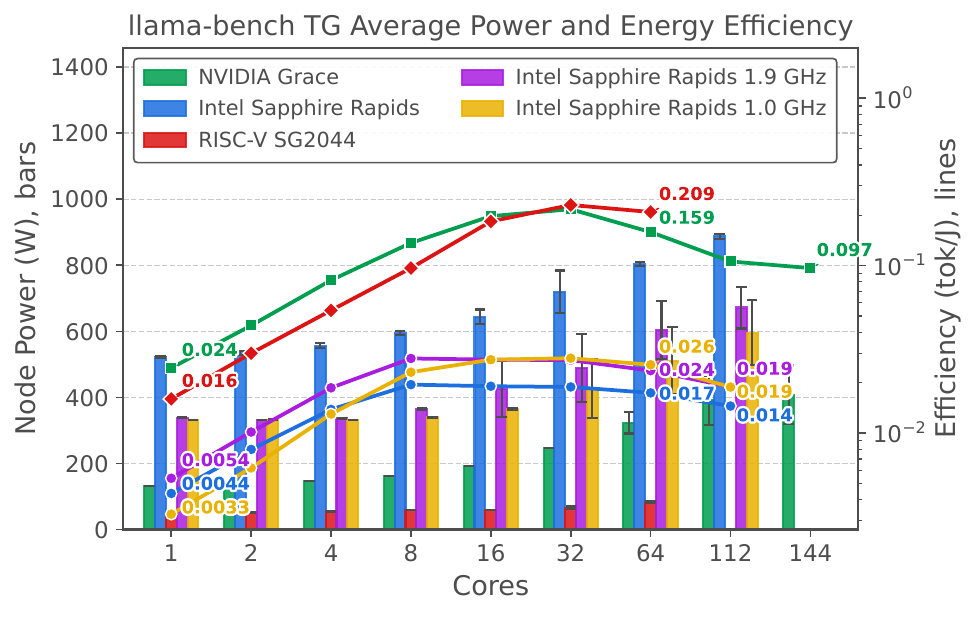}
            \caption{llama-bench Text-Generation}
            \label{fig:llama_power_efficiency_tg}
        \end{subfigure}
    \end{tabular}
    \caption{Synthetic benchmarks and application average power (bars, right axis) and energy efficiency (lines, left axis, HPL, HPCG, HPL-MxP, llama-bench) or energy to solution (lines, left axis, iPIC3D, PLUTO, OpenFOAM).}
    \label{fig:energy_efficiency}
\end{figure*}
In this section we continue the discussion of the results, evaluating the energy efficiency for each test. 
In general the plots, at the different tested core counts, report the average power (bars) and the respective energy efficiency (lines) accordingly to the performance metric of the benchmarks or applications, as discussed in the introduction of this section. 

As a general trend, in every test each machine shows similar average power and scaling behavior, suggesting that most of the power consumption is independent from the type of workload. In this context, the peak average power measured is in the neighborhood of 1000 W per the Intel Sapphire Rapids, 500 W for the NVIDIA Grace and 95 W for the Sophgo SG2044.
Moving to energy efficiency, the SG2044 achieves better values than the Intel Sapphire Rapids platform for several workloads, including HPL-MxP (Figure \ref{fig:hpl_mxp_power_efficiency}), OpenFOAM (Figure \ref{fig:openfoam_power_energy}), and llama-bench TG (Figure \ref{fig:llama_power_efficiency_tg}). In HPL (Figure \ref{fig:hpl_power_efficiency}) and iPIC3D (Figure \ref{fig:ipic3d_power_energy}), the two platforms exhibit similar efficiency, while HPCG (Figure \ref{fig:hpcg_power_efficiency}), PLUTO (Figure \ref{fig:pluto_power_energy}), and llama-bench PP (Figure \ref{fig:llama_power_efficiency_pp}) show a different trend, with comparable efficiency at lower core counts and a progressively higher efficiency for the SG2044 as the core count increases. These trends are correlated with the observed performance scaling. When the SG2044 sustains a near-ideal relative speedup, its energy efficiency increases with the number of active cores, progressively reducing the gap with Sapphire Rapids. Conversely, in workloads where Sapphire Rapids exhibits less favorable scaling, its increasing power consumption is not compensated by a proportional performance increase, resulting in lower energy efficiency at higher core counts. 
Compared with NVIDIA Grace, the SG2044 generally achieves lower energy efficiency, with two notable cases. In HPL, Sapphire Rapids achieves 4.21~GFLOP/J, compared with 3.98~GFLOP/J for NVIDIA Grace and 3.16~GFLOP/J for the SG2044 (Figure~\ref{fig:hpl_power_efficiency}). In llama-bench, instead, the SG2044 achieves the highest efficiency at 64 cores for both inference phases. For PP, it reaches 1.20~tok/J (Figure~\ref{fig:llama_power_efficiency_pp}), while for TG it reaches 0.209~tok/J (Figure~\ref{fig:llama_power_efficiency_tg}), against 1.14 and 0.159 tok/s of NVIDIA Grace, respectively. These results further illustrate how the relationship between performance scaling and power consumption can substantially affect the relative energy efficiency of the three architectures, particularly as the number of active cores increases. 

Considering the evolution of the RISC-V platforms across the Monte Cimone generations, the Sophgo SG2044 increases power consumption compared with the first generation SiFive U740, while also providing substantially higher energy efficiency. As shown in Figure~\ref{fig:hpl_power_efficiency}, a single MCv1 node, equipped with four SiFive U740 cores, achieves 0.315~GFLOP/J, compared with 0.533~GFLOP/J for the SG2044 in MCv3 at the corresponding data point. The latter is also close to the 0.537~GFLOP/J achieved by the Intel Sapphire Rapids platform. When comparing the aggregated node capabilities of the first and third Monte Cimone generations, energy efficiency increases from 0.315 to 3.16~GFLOP/J, corresponding to a $10\times$ improvement. At the same time, FP64 performance increases from 1.86 to 294~GFLOP/s (Figure~\ref{fig:hpl_raw_sub}), corresponding to a $158\times$ improvement.% These results quantify the substantial evolution in both computational performance and energy efficiency of the RISC-V platforms across the Monte Cimone generations.
\subsubsection{Frequency capping}
To ensure a fair comparison with the Intel Sapphire Rapids processor, given their substantially different base and maximum frequency we also conducted an evaluation with both a Memory Bound and a Compute Bound test while capping the maximum CPU frequency.
Figures~\ref{fig:llama_power_efficiency_pp} and~\ref{fig:llama_power_efficiency_tg} include two additional lines corresponding to these experiments. For PP, reducing the maximum frequency decreases energy efficiency, as this predominantly compute-bound workload experiences a reduction in throughput that is not sufficiently compensated by the lower power consumption. In contrast, for TG, which is primarily memory-bound, reducing the CPU frequency has little impact on throughput, particularly at 1.9~GHz. The reduction in power consumption therefore improves energy efficiency, although it remains below that achieved by the RISC-V and Arm platforms.

%% file: conclusions.tex
\section{Conclusions}\label{sec:conclusions}

In this work, we provided an extensive evaluation of the Sophgo SG2044 RISC-V processor, comparing it with two HPC-oriented platforms based on the other major processor ISAs: the Intel Xeon Platinum 8480+ based on x86-64 and the NVIDIA Grace CPU Superchip based on Arm.
We evaluated the SG2044 both as a standalone platform and as the compute node of Monte Cimone, a RISC-V cluster that has evolved through multiple generations of RISC-V hardware, providing a broader perspective on the development of RISC-V platforms for HPC.
In absolute performance, the SG2044 remains 8--15$\times$ behind the evaluated Intel and NVIDIA platforms across classical HPC workloads. However, it substantially improves computational capability and memory bandwidth over previous RISC-V generations, while its balance of per-core performance and system resources enables competitive scaling in HPCG, PLUTO, OpenFOAM, and LLM inference. Its energy efficiency is comparable to Intel across several workloads, while it exceeds Grace in the 64-core llama-bench tests and markedly improves over previous Monte Cimone generations.

Overall, these results indicate that RISC-V has progressed substantially from the performance levels of earlier HPC-oriented implementations and is increasingly capable of providing competitive system-level characteristics. The SG2044 scaling and energy-efficiency characteristics demonstrate that performance is not determined solely by peak computational capability. Future RISC-V HPC platforms can benefit from increased compute throughput improving vector capabilities, advancing its memory subsystem to achieve higher aggregated bandwidth, better adopted technology node to improve general efficiency and speed, and continued optimization of software stacks and numerical libraries. The evolution observed across the Monte Cimone generations suggests that these developments can progressively reduce the performance gap with established HPC architectures while preserving the flexibility of an open ISA.